\documentclass[12pt]{article}
\usepackage[T1]{fontenc}
\usepackage[utf8]{inputenc}
\usepackage{lmodern}
\usepackage{microtype}
\usepackage{amsmath,amssymb,amsthm,mathtools}
\usepackage{graphicx}
\usepackage[margin=1in]{geometry}
\usepackage{booktabs}
\usepackage{array}
\usepackage{xcolor}
\usepackage[font=small,labelfont=bf,margin=0.5cm]{caption}
\usepackage[authoryear,round]{natbib}
\usepackage{listings}
\usepackage[section]{placeins}
\usepackage[colorlinks=true,linkcolor={blue!45!black},citecolor={blue!45!black},urlcolor={blue!55!black}]{hyperref}

\definecolor{tabshade}{RGB}{238,243,249}
\newcommand{\E}{\mathbb{E}}
\newcommand{\R}{\mathbb{R}}
\newcommand{\C}{\mathbb{C}}
\newcommand{\N}{\mathcal{N}}
\newcommand{\KL}{\mathrm{KL}}
\newcommand{\tr}{\operatorname{tr}}
\newcommand{\Var}{\operatorname{Var}}

\newcommand{\dd}{\mathrm{d}}
\newcommand{\one}{\mathbf{1}}

\DeclareMathOperator{\Imag}{Im}
\DeclareMathOperator{\Beta}{Beta}

\theoremstyle{plain}
\newtheorem*{proposition*}{Proposition}
\newtheorem*{theorem*}{Theorem}
\newtheorem*{corollary*}{Corollary}
\newtheorem*{lemma*}{Lemma}
\theoremstyle{remark}
\newtheorem*{remark*}{Remark}

\lstdefinelanguage{lean}{
  keywords={theorem,def,abbrev,namespace,end,by,fun,let,in,match,with,where,induction,cases,intro,have,show,rw,simp,grind,decide,omega,exact,apply,unfold,constructor},
  sensitive=true, comment=[l]{--}, morecomment=[s]{/-}{-/}, string=[b]"}
\newcommand{\NAssets}{49}
\newcommand{\NDays}{9{,}254}
\newcommand{\SampleStart}{2 January 1990}
\newcommand{\SampleEnd}{30 September 2026}
\newcommand{\SnapQ}{0.196}

\newcommand{\SnapSigma}{0.502}
\newcommand{\SnapEdge}{1.046}
\newcommand{\SnapLambdaOne}{24.38}
\newcommand{\SnapNOut}{5}
\newcommand{\AbsMean}{0.545}
\newcommand{\AbsMax}{0.820}
\newcommand{\AbsMaxDate}{June 2012}
\newcommand{\AbsMin}{0.297}
\newcommand{\AbsMinDate}{September 2026}
\newcommand{\OutMedian}{6}
\newcommand{\OutMin}{4}
\newcommand{\OutMax}{9}
\newcommand{\OverlapMin}{0.798}
\newcommand{\NWindows}{429}
\newcommand{\CorrAbsOut}{0.41}
\newcommand{\MIp}{8}
\newcommand{\MIr}{4}
\newcommand{\MIn}{249}
\newcommand{\MIBias}{0.0660}
\newcommand{\MISd}{0.0165}
\newcommand{\MIPlugMean}{0.894}
\newcommand{\MICorrMean}{0.828}
\newcommand{\MINullPlugMean}{0.0653}
\newcommand{\MINullCorrMean}{\ensuremath{-}0.0007}
\newcommand{\MINullSd}{0.0171}
\newcommand{\MINullRej}{6.1}
\newcommand{\MIShareSig}{100.0}
\newcommand{\MICorrMax}{1.638}
\newcommand{\MICorrMaxDate}{April 2020}
\newcommand{\MICorrMin}{0.284}
\newcommand{\MICorrMinDate}{January 1996}
\newcommand{\MIPlugNoiseEight}{1.028}
\newcommand{\MICorrNoiseEight}{0.827}
\newcommand{\MIBiasNoiseEight}{0.201}
\newcommand{\MIPlugNoiseTwentyFour}{1.319}
\newcommand{\MICorrNoiseTwentyFour}{0.833}
\newcommand{\MIBiasNoiseTwentyFour}{0.487}
\newcommand{\MIPartialMean}{0.170}
\newcommand{\MIPartialBias}{0.0663}
\newcommand{\MIPartialShareSig}{100.0}
\newcommand{\MIPartialMax}{0.490}
\newcommand{\MIPartialMaxDate}{October 2008}
\newcommand{\MIPartialMin}{0.034}
\newcommand{\MIPartialMinDate}{February 1996}
\newcommand{\MIPartialAbs}{0.05}
\newcommand{\MIPartialShare}{18.6}
\newcommand{\MICorrAbs}{0.82}
\newcommand{\KLfwdN}{3.515}
\newcommand{\KLrevN}{2.641}
\newcommand{\KLfwdSixty}{120.05}
\newcommand{\KLrevSixty}{16.07}
\newcommand{\CCAEdge}{0.500}
\newcommand{\CCAThr}{0.408}

\newcommand{\FrontierSim}{0.0912}
\newcommand{\FrontierEmp}{0.0118}
\newcommand{\FrontierEmpT}{0.153}
\newcommand{\CCAsixA}{0.5748}
\newcommand{\CCAsixS}{0.5732}
\newcommand{\CCAeightA}{0.7521}
\newcommand{\CCAeightS}{0.7524}
\newcommand{\MOUrate}{6.88}
\newcommand{\EWVol}{17.93}
\newcommand{\VolSampleSixty}{23.51}
\newcommand{\VolClipSixty}{11.87}
\newcommand{\VolLWSixty}{11.99}
\newcommand{\VolRIESixty}{15.37}

\newcommand{\RatioTwoFifty}{2.06}
\newcommand{\TheoryTwoFifty}{1.55}
\newcommand{\VolSampleFiveHundred}{11.59}
\newcommand{\VolClipFiveHundred}{11.93}

\newcommand{\RatioFiveHundred}{1.63}
\newcommand{\TheoryFiveHundred}{1.23}

\newcommand{\MCRepsQuarter}{1{,}000}
\newcommand{\MCReps}{4{,}000}

\title{Spurious Information in Large Markets:\\[0.3em]
\large Random Matrix Laws for Entropy, Mutual Information and the Value of Information}
\author{James K Rice\thanks{PhD Candidate, University of Essex; \href{mailto:james.k.rice@essex.ac.uk}{james.k.rice@essex.ac.uk}. Replication materials, including the Lean formalization, are available at \url{https://github.com/jameskrice7/RMT}.}}
\date{November 2026}

\begin{document}
\maketitle

\begin{abstract}
\noindent Entropy, Kullback--Leibler divergence and mutual information have become standard currencies for measuring how much one set of financial variables reveals about another. In large cross-sections they are almost always computed from sample covariance matrices, and the noise in those matrices is the object of random matrix theory. This paper develops the random matrix theory of Gaussian information functionals and applies it to markets. I show that the plug-in mutual information between two blocks of $p$ and $r$ variables estimated from $n$ observations has an exact expected excess, the \emph{spurious information}, that is a digamma telescope independent of the true dependence. Under independence its law is that of a product of Beta variables. In the proportional regime it grows linearly in $n$ at the rate $\kappa(c_1,c_2)=\tfrac12[(1-c_1-c_2)\log(1-c_1-c_2)-(1-c_1)\log(1-c_1)-(1-c_2)\log(1-c_2)]$ with an order-one variance. The same quantity is a linear spectral statistic of the Jacobi ensemble, which yields a closed form for the logarithmic moment of Wachter's law. Spurious information increases strictly whenever uninformative coordinates are added. In a CARA--Gaussian economy the certainty-equivalent value of a signal equals mutual information divided by risk aversion, so these results translate one for one into a spurious value of information. The canonical-correlation phase transition then defines a frontier of minimum detectable information. For correlation dynamics I show that the Dyson eigenvalue process of a matrix Ornstein--Uhlenbeck market is informationally sufficient for its mean-reversion rate and carries an information rate of order $N^2$. Monte Carlo experiments confirm every exact and limiting formula. On daily returns of 49 US industry portfolios from 1990 to 2026, I document the market mode, the reliability of sector eigenvectors implied by the BBP transition, the out-of-sample cost of noise dressing, and an information linkage between carbon-intensive and financial industries that survives exact bias correction and conditioning on the market. Selected algebraic and combinatorial steps are machine-checked in Lean~4.
\end{abstract}




\section*{Introduction}

Information is the commodity that financial markets process, and over the last two decades its quantitative study has moved from verbal description to measurement. Entropy and mutual information appear in models of rational inattention \citep{sims2003implications,matejka2015rational,mackowiak2023rational}, in theories of information acquisition and portfolio choice \citep{vannieuwerburgh2010information,veldkamp2011information,kacperczyk2016rational}, in the ordering of signals by their value to investors \citep{cabrales2013entropy,kadan2019estimating}, and in a large empirical literature that measures connectedness, spillovers and the informational content of one market for another \citep{billio2012econometric,diebold2014network}. The Kullback--Leibler divergence \citep{kullback1951information} and mutual information \citep{shannon1948mathematical,cover2006elements} have become the natural units in which such quantities are reported. Whenever the variables are numerous, these units are computed from covariance matrices, typically through the Gaussian formula $I=\tfrac12\log[\det\Sigma_{11}\det\Sigma_{22}/\det\Sigma]$ or through closely related log-determinant expressions. The covariance matrices are themselves estimated, and in modern financial applications the number of variables is rarely small compared with the number of observations.

Random matrix theory describes what happens to such estimates when dimension and sample size are of the same order. Since the work of \citet{laloux1999noise} and \citet{plerou1999universal}, it has been a central tool of econophysics. Those papers showed that the bulk of the eigenvalues of an empirical correlation matrix of stock returns is close to the Marchenko--Pastur law of pure noise \citep{marchenko1967distribution}, that a handful of eigenvalues escape the bulk, and that the largest of them describes a market mode. Subsequent work established the statistical theory of these outliers \citep{johnstone2001distribution,baik2005phase,paul2007asymptotics,benaych2011eigenvalues}, developed estimators that clean the noise from the spectrum \citep{ledoit2004well,ledoit2012nonlinear,bun2016rotational,bun2017cleaning,ledoit2020analytical}, and quantified the risk underestimation of optimised portfolios \citep{pafka2003noisy,elkaroui2010high,bodnar2018estimation}. The monographs of \citet{bai2010spectral}, \citet{anderson2010introduction} and \citet{potters2020first} collect the mathematics. Yet the random matrix theory of \emph{information functionals} of sample covariance matrices, as distinct from their spectra, eigenvectors or portfolio risk, has received far less attention in finance. That gap matters because the information functionals are exactly the quantities that economic theory gives a price to.

This paper develops that theory and applies it. The organising object is the plug-in Gaussian mutual information between two blocks of variables, of dimensions $p$ and $r$, estimated from $n$ observations. I show that its expected excess over the truth, which I call the \emph{spurious information}, is an exact finite-sample function of $(p,r,n)$ alone and is unaffected by the true covariance matrix. It is a telescoping sum of digamma differences, so that its positivity and its strict monotonicity in each block dimension are elementary consequences of the monotonicity of the digamma function. Under independence of the blocks, the plug-in information is minus one half the logarithm of Wilks' statistic. Its exact law is therefore that of a sum of independent log-Beta variables, which gives the exact null variance and an exact test. In the proportional regime $p/n\to c_1$, $r/n\to c_2$, the spurious information grows linearly in $n$ at the rate $\kappa(c_1,c_2)$ displayed in the abstract, while its standard deviation stays of order one. A naive count of parameters would predict the classical value $pr/(2n)$, which underestimates the spurious information by a factor that is already close to two when $c_1=c_2=0.4$.

The same quantity has a second, purely spectral description. The plug-in information equals $-\tfrac12\sum_k\log(1-\hat\rho_k^2)$, where $\hat\rho_k$ are the sample canonical correlations. Under independence their empirical distribution converges to the law of \citet{wachter1980limiting}, the limiting spectral law of the Jacobi or MANOVA ensemble \citep{johnstone2008multivariate,collins2005product}. Matching the two descriptions yields a closed-form evaluation of the logarithmic moment of the Wachter law, and the same argument identifies the variance of the corresponding linear spectral statistic. I also use the finite-rank theory of canonical correlations \citep{bao2019canonical} to describe a phase transition for information. A single dependence direction with population canonical correlation $t$ produces a separated sample canonical correlation only if $t^4>c_1c_2/[(1-c_1)(1-c_2)]$. The corresponding mutual information, $I_c(c_1,c_2)=-\tfrac12\log(1-t_c^2)$, is a frontier below which information exists but cannot be located by spectral methods.

These results translate directly into economics. In a price-taking CARA--Gaussian portfolio problem, I show that the ratio of ex-ante expected utilities with and without a jointly Gaussian signal is exactly $\exp\{-I(v;s)\}$, where $v$ is the vector of payoffs and $s$ the signal. The certainty-equivalent value of information is therefore $I(v;s)/\gamma$, mutual information measured in units of risk tolerance. The identity is in the spirit of \citet{grossman1980impossibility}, \citet{vannieuwerburgh2010information} and \citet{cabrales2013entropy}, but here it holds as an exact equality for arbitrary dimensions. It implies that any econometrician, investor or data vendor who estimates the value of a high-dimensional signal set from data inherits the spurious information as a \emph{spurious value of information}, with the same exact expectation, null law and proportional limit. Because the spurious value grows strictly when uninformative series are appended, a seller can inflate the measured value of a signal set without adding information. A buyer can remove the inflation exactly with a digamma correction. The same exact theory yields the expected Kullback--Leibler cost of acting on plug-in Gaussian beliefs in either direction. These costs are Stein's loss and the entropy loss in information-theoretic form \citep{james1961estimation,dey1985estimation}, and I give their finite-sample values and high-dimensional limits.

Financial correlations are not static, and econophysics has emphasised their regime-like dynamics \citep{munnix2012identifying,schmitt2013non,allez2012eigenvector}. I therefore study the information contained in a continuously observed matrix Ornstein--Uhlenbeck process, whose eigenvalues perform a Dyson Brownian motion with confinement \citep{dyson1962brownian,chan1992wigner,rogers1993interacting}. Its Stieltjes transform obeys a forced complex Burgers equation whose stationary solution is the semicircle. Girsanov's theorem gives the Kullback--Leibler rate between two mean-reversion regimes as $N(N+1)(b-a)^2/(8a)$. The same rate is obtained from the eigenvalue process alone, so the Dyson process is informationally sufficient for the regime and the eigenvector motion carries no information about it. By the Chernoff--Stein lemma, the time needed to tell two regimes apart therefore falls like $N^{-2}$. This sufficiency result connects to variational inference for latent stochastic differential equations \citep{li2020scalable,rice2026sldi}, where the path-space Kullback--Leibler divergence is the regulariser. For rotation-equivariant drift families, inference on the eigenvalue process loses nothing.

The empirical application uses \NDays{} daily value-weighted returns of the \NAssets{} Fama--French industry portfolios from \SampleStart{} to \SampleEnd{} \citep{fama1997industry,french2026data}. The data reproduce the familiar econophysics picture of a Marchenko--Pastur bulk with a dominant market mode and a few sector modes. Inverting the BBP outlier map lets me attach to each sector eigenvector a predicted overlap with its population counterpart, which distinguishes the reliable modes from those that are barely above the noise. Over \NWindows{} rolling windows the market mode absorbs between \AbsMin{} and \AbsMax{} of total variance. A minimum-variance backtest exhibits the noise-dressing penalty predicted by random matrix theory and the benefit of cleaning, which is large when the estimation window is short. The information analysis asks how much the returns of eight carbon-intensive industries reveal about four financial industries, a question motivated by the climate-finance literature on transition risk \citep{battiston2017climate,bolton2021investors,giglio2021climate}. The exact spurious information for this design is \MIBias{} nats with a null standard deviation of \MISd{}. In a time-shifted placebo, the plug-in information averages \MINullPlugMean{} nats, the corrected placebo averages \MINullCorrMean{}, and the corrected test rejects in \MINullRej\% of windows at the nominal 5\% level. The genuine linkage averages \MICorrMean{} nats after correction and is strongly associated with the market mode. Conditioning on the market leaves a smaller linkage of \MIPartialMean{} nats on average. It is significant in every window and uncorrelated with the absorption ratio, and it peaks in \MIPartialMaxDate{}.

The paper draws on, and speaks to, several literatures. Within random matrix theory it builds on the classical theory of Gaussian and Wishart ensembles \citep{mehta2004random,forrester2010log,dumitriu2002matrix,edelman2005random}, on the local laws and anisotropic estimates that make the outlier theory uniform \citep{knowles2017anisotropic}, and on the handbook and textbook treatments of \citet{akemann2011oxford}, \citet{livan2018introduction} and \citet{couillet2022random}. The latter emphasises, as this paper does, that random matrix corrections are first order in modern data analysis. Optimal eigenvalue shrinkage and thresholding in spiked models \citep{gavish2014optimal,donoho2018optimal}, regularisation by banding and thresholding \citep{bickel2008regularized}, factor-based covariance estimation \citep{fan2013large}, and dynamic covariance models with nonlinear shrinkage \citep{engle2002dynamic,engle2019large} form the estimation toolkit against which the information-specific corrections here should be compared. \citet{ledoit2022power} survey that toolkit, and its portfolio applications go back to \citet{ledoit2003improved} and \citet{ledoit2017nonlinear}. In econophysics, the random matrix analysis of financial correlations was extended by \citet{gopikrishnan2001quantifying}, \citet{plerou2002random}, \citet{utsugi2004random} and \citet{burda2005spectral}, and was complemented by hierarchical and network descriptions \citep{mantegna1999hierarchical,tumminello2010correlation}. Textbook accounts include \citet{mantegna2000introduction}, \citet{bouchaud2003theory} and the review of \citet{kwapien2012physical}. The portfolio consequences of estimation noise, from \citet{markowitz1952portfolio} through \citet{elton1973estimating}, \citet{michaud1989markowitz}, \citet{jagannathan2003risk}, \citet{pafka2002noisy} and \citet{kondor2007noise}, motivate the minimum-variance exercise. In financial econometrics, the determination of the number of factors \citep{bai2002determining,onatski2009testing,onatski2010determining,ahn2013eigenvalue} is itself a random matrix problem rooted in the arbitrage pricing theory \citep{ross1976arbitrage,chamberlain1983arbitrage}. So is the estimation of latent factors from large panels and high-frequency data \citep{lettau2020factors,lettau2020estimating,pelger2020understanding,aitsahalia2017using}. Random matrix asymptotics also explain the surprising behaviour of overparameterised regression \citep{hastie2022surprises}. In information theory, random matrix methods have long been used to compute the capacity of multi-antenna channels, which is a log-determinant functional of exactly the kind studied here \citep{telatar1999capacity,tulino2004random}. The difficulty of estimating mutual information in high dimensions is well documented for nonparametric and neural estimators \citep{paninski2003estimation,kraskov2004estimating,belghazi2018mutual,mcallester2020formal}. The Gaussian case studied here is the one in which that difficulty can be computed exactly. In economics, the comparison of information structures goes back to \citet{blackwell1953equivalent}. The role of information in markets is surveyed by \citet{stiglitz2000contributions}, and its incorporation into prices is the subject of the efficiency and microstructure literatures \citep{fama1970efficient,kyle1985continuous}.

The paper makes four contributions. The first is a set of exact and asymptotic laws for spurious information, Kullback--Leibler costs and their fluctuations, derived from the Bartlett decomposition, the Jacobi ensemble and the Marchenko--Pastur and Wachter laws, including a new closed form for the logarithmic moment of the Wachter law. The second is the economic translation of these laws through an exact value-of-information identity, together with the monotonicity, correction and frontier results it implies. The third is an information-theoretic analysis of Dyson dynamics that establishes the sufficiency of the eigenvalue process and the quadratic scaling of regime detectability. The fourth is an econophysics application that combines these tools on more than three decades of data. Every formula is checked against simulation, and the combinatorial and algebraic steps that link the analytic arguments are verified in the Lean~4 theorem prover \citep{demoura2021lean}.

The paper proceeds as follows. The next section reviews the random matrix machinery used throughout, namely resolvents, Stieltjes and $S$-transforms, the Marchenko--Pastur and Wachter laws, spiked models and Dyson Brownian motion. I then derive the information laws and their economic interpretation, followed by the dynamic results. A numerical section verifies the theory, and the application section takes it to the data. The appendices contain proofs, a description of the Lean formalization and additional tables.
\section*{Random Matrix Preliminaries}

Let $X\in\R^{n\times N}$ have independent rows $x_t^\top\sim\N(0,\Sigma)$, $t=1,\ldots,n$, so that $W=X^\top X$ has the Wishart law $\mathcal W_N(n,\Sigma)$ and $S=W/n$ is the sample covariance matrix with known zero mean. When the mean is estimated, all statements hold with $n$ replaced by $n-1$, since the centred cross-product matrix is $\mathcal W_N(n-1,\Sigma)$ \citep[Ch.~3]{muirhead1982aspects}. The aspect ratio is $q=N/n$. Throughout, $\lambda_1\geq\cdots\geq\lambda_N$ denote the eigenvalues of $S$ and $\mu_S=N^{-1}\sum_i\delta_{\lambda_i}$ its empirical spectral measure. This section collects the tools of random matrix theory used later. Its purpose is to fix conventions and to give self-contained derivations of the few facts on which the new results rest. Comprehensive treatments are \citet{bai2010spectral}, \citet{anderson2010introduction}, \citet{tao2012topics}, \citet{erdos2017dynamical} and, with an eye to applications in finance, \citet{bouchaud2011financial}, \citet{bun2017cleaning} and \citet{potters2020first}.

\subsection*{Resolvents, Stieltjes transforms and the Marchenko--Pastur law}

For $z\in\C\setminus\R$ the resolvent is $G_S(z)=(z-S)^{-1}$ and the Stieltjes transform of $\mu_S$ is
\[
 g_S(z)=\frac1N\tr G_S(z)=\int\frac{\mu_S(\dd x)}{z-x}.
\]
The function $g_S$ is analytic off the spectrum, maps the upper half-plane to the lower half-plane, and behaves like $1/z$ at infinity, with the Laurent expansion $g_S(z)=\sum_{k\geq0}m_k z^{-k-1}$ in terms of the moments $m_k=N^{-1}\tr S^k$. A probability density is recovered from its Stieltjes transform through the inversion formula $\rho(x)=\pi^{-1}\lim_{\eta\downarrow0}\Imag g(x-i\eta)$. Convergence of Stieltjes transforms at every $z\in\C\setminus\R$ is equivalent to weak convergence of the underlying measures. The entire asymptotic theory of spectra can therefore be phrased in terms of $g$.

\begin{figure}[htbp]
\centering
\includegraphics[width=0.62\textwidth]{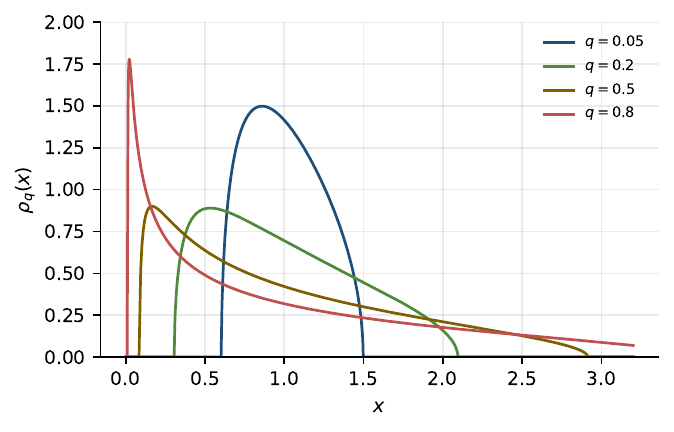}
\caption{Marchenko--Pastur densities $\rho_q$ for four aspect ratios. The edges $(1\pm\sqrt q)^2$ are the zeros of the discriminant of the quadratic \eqref{eq:mp-quadratic}. The left edge reaches the origin at $q=1$, where $\log\det S$ diverges.}
\label{fig:mp-family}
\end{figure}

The self-consistent equation for $g_S$ follows from a leave-one-out argument that I record because the same algebra reappears in the outlier and cleaning formulas. Write $S=n^{-1}\sum_t x_tx_t^\top$ and $S_{(t)}=S-n^{-1}x_tx_t^\top$, with resolvent $G_{(t)}=(z-S_{(t)})^{-1}$. Taking the trace of the identity $zG_S-SG_S=I$ gives $zNg_S-n^{-1}\sum_t x_t^\top G_Sx_t=N$. The Sherman--Morrison formula gives $G_S=G_{(t)}+G_{(t)}x_tx_t^\top G_{(t)}/(n-x_t^\top G_{(t)}x_t)$, whence
\[
 \frac1n x_t^\top G_Sx_t=\frac{n^{-1}x_t^\top G_{(t)}x_t}{1-n^{-1}x_t^\top G_{(t)}x_t}.
\]
Because $x_t$ is independent of $G_{(t)}$, the quadratic form $n^{-1}x_t^\top G_{(t)}x_t$ concentrates around $n^{-1}\tr(\Sigma G_{(t)})$, and $G_{(t)}$ differs from $G_S$ by a rank-one perturbation whose normalised trace is $O(N^{-1})$. When $\Sigma=I$ the quadratic form is therefore $qg_S+o(1)$, and dividing the trace identity by $N$ yields $zg-g/(1-qg)=1$. In polynomial form,
\begin{equation}
 qz\,g(z)^2-(z-1+q)\,g(z)+1=0.
 \label{eq:mp-quadratic}
\end{equation}
The root with $g(z)\sim1/z$ at infinity is $g(z)=[z-1+q-\sqrt{(z-1+q)^2-4qz}\,]/(2qz)$. The discriminant vanishes at $\lambda_\pm=(1\pm\sqrt q)^2$, and the inversion formula yields the Marchenko--Pastur density
\begin{equation}
 \rho_q(x)=\frac{\sqrt{(\lambda_+-x)(x-\lambda_-)}}{2\pi qx},\qquad x\in[\lambda_-,\lambda_+],
 \label{eq:mp-density}
\end{equation}
together with an atom of mass $1-1/q$ at the origin when $q>1$ \citep{marchenko1967distribution}. The edges are also the almost sure limits of the extreme eigenvalues \citep{geman1980limit,bai1993limit}. Figure~\ref{fig:mp-family} shows the family. As $q$ grows the bulk spreads from a point mass at one to an interval that touches the origin at $q=1$. This spreading is the mechanism behind every cost of dimensionality discussed in this paper.

The moments of $\rho_q$ follow from \eqref{eq:mp-quadratic} without integration. Writing $h(x)=\sum_k m_kx^k=z\,g(z)$ with $x=1/z$, the quadratic becomes $h=1+xh(1-q+qh)$, which is equivalent to the recursion
\begin{equation}
 m_k=(1-q)\,m_{k-1}+q\sum_{j=0}^{k-1}m_j\,m_{k-1-j},\qquad m_0=1.
 \label{eq:mp-recursion}
\end{equation}
Its solution is the Narayana polynomial $m_k=\sum_{j=0}^{k-1}\tfrac1k\binom kj\binom k{j+1}q^j$, whose coefficients count non-crossing partitions of $\{1,\ldots,k\}$ by number of blocks \citep{stanley1999enumerative,nica2006lectures}. The first four are $1$, $1+q$, $1+3q+q^2$ and $1+6q+6q^2+q^3$. At $q=1$ the Narayana numbers sum to the Catalan numbers, which are also the even moments of the semicircle law $\rho_{\rm sc}(x)=(2\pi)^{-1}\sqrt{4-x^2}$ of \citet{wigner1955characteristic,wigner1958distribution}. The Lean file accompanying this paper verifies with exact integer arithmetic that the recursion \eqref{eq:mp-recursion} generates the Narayana polynomials up to order ten, that the semicircle recursion generates the Catalan numbers up to order twelve, and that the Narayana rows sum to the Catalan numbers. At finite $n$ the exact Wishart moments differ from the limits by explicit corrections. Using $\E W^2=n(n+N+1)I$ and $\E\tr W^3=nN(n^2+N^2+3nN+3n+3N+4)$ for $\Sigma=I$ \citep{muirhead1982aspects},
\[
 \frac1N\E\tr S^2=1+q+\frac1n,\qquad
 \frac1N\E\tr S^3=1+3q+q^2+\frac{3(1+q)}n+\frac4{n^2}.
\]
Both formulas are checked by simulation in the numerical section.

For general $\Sigma$ with limiting spectral distribution $H$ the same leave-one-out argument, now keeping track of $n^{-1}\tr(\Sigma G_{(t)})$, gives the Marchenko--Pastur--Silverstein equation
\begin{equation}
 g(z)=\int\frac{H(\dd\tau)}{z-\tau\,w(z)},\qquad w(z)=1-q+qz\,g(z),
 \label{eq:silverstein}
\end{equation}
\citep{silverstein1995empirical}. The function $w$ is a subordination function: the sample resolvent evaluated at $z$ behaves on average like the population resolvent evaluated at $z/w(z)$, rescaled. Three facts used below are direct consequences of \eqref{eq:silverstein}. First, for $H=\delta_1$ it reduces to $g=1/(z-w)$, which is \eqref{eq:mp-quadratic}. Second, a population eigenvalue $\ell$ outside the bulk produces a sample outlier at the solution $z$ of $z=\ell\,w(z)$, the value at which the integrand of \eqref{eq:silverstein} becomes singular. Third, the expected projection of a sample eigenvector on the population eigenbasis is a function of $w$ evaluated at the sample eigenvalue. That fact is the basis of the oracle nonlinear shrinkage formula
\begin{equation}
 \xi(\lambda)=\frac{\lambda}{|1-q+q\lambda\,g(\lambda-i0)|^2}=\frac{\lambda}{|w(\lambda-i0)|^2}
 \label{eq:rie}
\end{equation}
of \citet{ledoit2011eigenvectors} and the rotationally invariant estimator of \citet{bun2016rotational}. In this estimator, each sample eigenvalue is replaced by $\xi(\lambda_i)$, with $g$ evaluated from the sample spectrum at a small distance $\eta$ below the real axis.

\subsection*{Free probability}

The Silverstein equation is an instance of free multiplicative convolution. For a probability measure $\mu$ on $[0,\infty)$ with moment generating function $\psi_\mu(z)=\sum_{k\geq1}m_kz^k$, let $\chi_\mu$ be the inverse of $\psi_\mu$ near the origin and define the $S$-transform $S_\mu(w)=\chi_\mu(w)(1+w)/w$. \citet{voiculescu1991limit} showed that large rotationally invariant matrices behave as free random variables, and that the spectrum of a product of free positive elements has $S$-transform equal to the product of the $S$-transforms \citep{voiculescu1992free,mingo2017free}. From \eqref{eq:mp-recursion}, $\psi=h-1$ satisfies $\psi=x(1+\psi)(1+q\psi)$, so $\chi(w)=w/[(1+w)(1+qw)]$ and
\[
 S_{\rm MP}(w)=\frac1{1+qw}.
\]
The spectrum of $S=\Sigma^{1/2}S_0\Sigma^{1/2}$, with $S_0$ white Wishart, is then the free multiplicative convolution $H\boxtimes\rho_q$, whose $S$-transform is $S_H(w)/(1+qw)$. Two consequences are worth stating. Because $S_{\rm MP}$ is a M\"obius map, cleaning a sample spectrum amounts to a free deconvolution, which is what \eqref{eq:rie} implements eigenvalue by eigenvalue. Free cumulants of $\rho_q$ are all equal to $q^{k-1}$, which is the free-probability statement of the Narayana formula. The additive analogue is the $R$-transform, $R_\mu=g_\mu^{-1}(w)-1/w$. The semicircle has $R(w)=\sigma^2w$, which drives the dynamics studied later.

\subsection*{Canonical correlations and the Jacobi ensemble}

\begin{figure}[htbp]
\centering
\includegraphics[width=\textwidth]{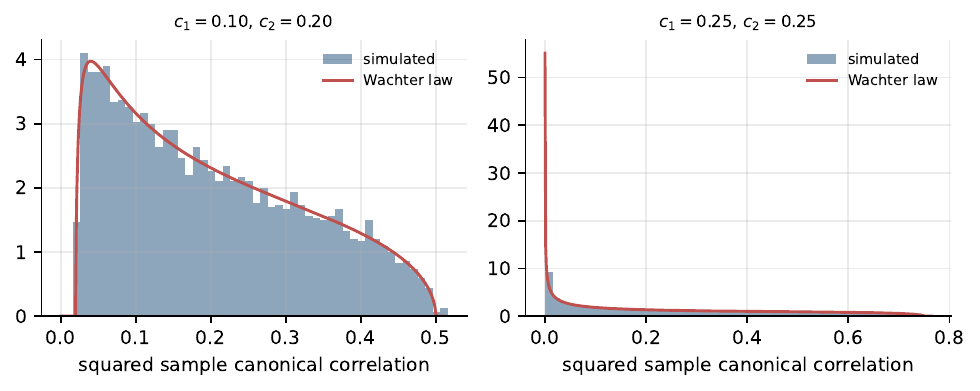}
\caption{Squared sample canonical correlations of independent Gaussian blocks (30 replications) and the Wachter law \eqref{eq:wachter}. Left: $n=1000$, $p=100$, $r=200$. Right: $n=600$, $p=r=150$. The upper edge $\omega_+$ stays below one because $c_1+c_2<1$.}
\label{fig:wachter}
\end{figure}

Partition $x_t=(x_{t,1}^\top,x_{t,2}^\top)^\top$ with blocks of dimensions $p\leq r$, and partition $S$ conformably. The squared sample canonical correlations $\hat\rho_1^2\geq\cdots\geq\hat\rho_p^2$ are the eigenvalues of $S_{11}^{-1}S_{12}S_{22}^{-1}S_{21}$. Equivalently, if $P_1$ and $P_2$ are the orthogonal projections of $\R^n$ onto the column spaces of the two data blocks, they are the nonzero eigenvalues of $P_1P_2P_1$. When the blocks are independent, the column spaces are uniformly distributed and independent. The $\hat\rho_k^2$ then form a Jacobi (MANOVA) ensemble, with joint density proportional to $\prod_k x_k^{(r-p-1)/2}(1-x_k)^{(n-r-p-1)/2}\prod_{j<k}|x_j-x_k|$ on $[0,1]^p$ \citep{muirhead1982aspects,johnstone2008multivariate}. In free-probability terms, $P_1$ and $P_2$ are asymptotically free projections of traces $c_1=p/n$ and $c_2=r/n$, and the limiting law of $P_1P_2P_1$ is the free multiplicative convolution of two Bernoulli laws \citep{collins2005product}. Its absolutely continuous part, for $c_1\le c_2$ and $c_1+c_2<1$, is the law of \citet{wachter1980limiting},
\begin{equation}
 w_{c_1,c_2}(x)=\frac{\sqrt{(x-\omega_-)(\omega_+-x)}}{2\pi c_1\,x(1-x)},\qquad
 \omega_\pm=\Bigl(\sqrt{c_1(1-c_2)}\pm\sqrt{c_2(1-c_1)}\Bigr)^2.
 \label{eq:wachter}
\end{equation}
This density has unit mass on $[\omega_-,\omega_+]$ and mean $c_2$. Figure~\ref{fig:wachter} compares it with simulated canonical correlations of independent Gaussian blocks. The upper edge satisfies $\omega_+<1$ exactly when $c_1+c_2<1$, which keeps $\log(1-x)$ bounded on the support. That boundedness drives the information laws below.

\subsection*{Spiked models and phase transitions}

Real covariance matrices are not white, and the structures that matter in finance are low-rank: a market factor, a handful of sectors, a few dependence directions between blocks. The spiked covariance model $\Sigma=I+\sum_{k\leq K}(\ell_k-1)v_kv_k^\top$ with fixed $K$ captures this. Substituting $H=(1-K/N)\delta_1+N^{-1}\sum_k\delta_{\ell_k}$ into \eqref{eq:silverstein} and using the outlier condition $z=\ell w(z)$ with the Marchenko--Pastur subordination function gives the explicit relation
\begin{equation}
 \lambda(\ell)=\ell\Bigl(1+\frac q{\ell-1}\Bigr),\qquad\ell>1+\sqrt q.
 \label{eq:bbp}
\end{equation}
The derivation is short. With $w=z/\ell$, the bulk relation $(w-1+q)(z-w)=qz$ becomes $(z/\ell-1+q)(1-1/\ell)=q$, which rearranges to \eqref{eq:bbp}. The solution leaves the bulk only when $\lambda(\ell)>\lambda_+$. Since $\lambda(1+\sqrt q)=(1+\sqrt q)^2$, the threshold is $\ell_c=1+\sqrt q$, the Baik--Ben Arous--P\'ech\'e (BBP) transition \citep{baik2005phase,baik2006eigenvalues}. Above the threshold the sample eigenvector $u_k$ makes a nontrivial angle with $v_k$,
\begin{equation}
 \langle u_k,v_k\rangle^2\to\frac{1-q/(\ell_k-1)^2}{1+q/(\ell_k-1)}.
 \label{eq:overlap}
\end{equation}
Below it, the overlap vanishes \citep{paul2007asymptotics,benaych2011eigenvalues,bloemendal2016principal}. Inverting \eqref{eq:bbp} gives the quadratic $\ell^2-(\lambda+1-q)\ell+\lambda=0$, whose larger root is the population spike implied by an observed outlier $\lambda$, and \eqref{eq:overlap} then predicts how reliable the corresponding eigenvector is. The Lean file certifies the threshold identity, the vanishing of the overlap numerator at the threshold and the inversion quadratic. Fluctuations of the largest eigenvalue at the edge follow the Tracy--Widom laws \citep{tracy1994level,tracy1996orthogonal,johnstone2001distribution}. They are not needed here.

The analogous transition for canonical correlations is less familiar and plays a central role in the information frontier below. Suppose the blocks have a single nonzero population canonical correlation $t$. The theory of \citet{bao2019canonical} shows that the leading sample canonical correlation separates from the Wachter bulk if and only if $t>t_c$, and otherwise sticks to $\omega_+$. In the parametrisation I use, which simulation confirms in the numerical section, the threshold and the separated limit are
\begin{equation}
 t_c^4=\frac{c_1c_2}{(1-c_1)(1-c_2)},\qquad
 \hat\rho_1^2\to\gamma(t)=t^2\Bigl(1-c_1+\frac{c_1}{t^2}\Bigr)\Bigl(1-c_2+\frac{c_2}{t^2}\Bigr)\quad(t>t_c).
 \label{eq:cca-spike}
\end{equation}
At $t=t_c$, $\gamma(t_c)=c_1(1-c_2)+c_2(1-c_1)+2t_c^2(1-c_1)(1-c_2)=\omega_+$, because $t_c^2(1-c_1)(1-c_2)=\sqrt{c_1c_2(1-c_1)(1-c_2)}$. At $t=1$, $\gamma=1$. The Lean file proves the threshold identity under the defining constraint on $t_c$. As with \eqref{eq:bbp}, the map $t\mapsto\gamma(t)$ is invertible above the threshold. A quadratic in $t^2$ recovers a debiased canonical correlation from an observed outlier.

\subsection*{Logarithmic spectral statistics}

The functionals at the heart of this paper are logarithmic. The Bartlett decomposition writes $W=LL^\top$ for $\Sigma=I$, with $L$ lower triangular, independent entries, $L_{ii}^2\sim\chi^2_{n-i+1}$ and $L_{ij}\sim\N(0,1)$ for $i>j$ \citep[Ch.~3]{muirhead1982aspects}. Hence $\log\det W=\sum_{i=1}^N\log L_{ii}^2$ is a sum of independent log-chi-square variables. Since $\E\log\chi^2_k=\psi(k/2)+\log2$ and $\Var\log\chi^2_k=\psi'(k/2)$, with $\psi$ the digamma function, and since $\det S=\det\Sigma\,\det(W_0/n)$ for a white Wishart $W_0$, one obtains for every $\Sigma$ the exact identities
\begin{equation}
 \begin{gathered}
 \ell_n(N)\equiv\E\log\det S-\log\det\Sigma=\sum_{i=1}^N\Bigl[\psi\Bigl(\frac{n-i+1}2\Bigr)-\log\frac n2\Bigr],\\
 \Var\log\det S=\sum_{i=1}^N\psi'\Bigl(\frac{n-i+1}2\Bigr).
 \end{gathered}
 \label{eq:logdet}
\end{equation}
Using $\psi(x)=\log x-1/(2x)+O(x^{-2})$ and $\psi'(x)=1/x+O(x^{-2})$, a Riemann-sum argument gives
\begin{equation}
 \frac1N\ell_n(N)\to\varphi(q)\equiv\int\log x\,\rho_q(\dd x)=-1-\frac{1-q}q\log(1-q),\qquad
 \Var\log\det S\to-2\log(1-q),
 \label{eq:phi}
\end{equation}
because $N^{-1}\sum_{i\le N}\log(1-(i-1)/n)\to q^{-1}\int_0^q\log(1-u)\,\dd u$. The second-order term is explicit as well. An Euler--Maclaurin expansion shows $\ell_n(N)=N\varphi(N/n)+\tfrac12\log(1-N/n)+o(1)$, which is the mean correction in the central limit theorem for linear spectral statistics of \citet{bai2004clt}, with variance $-2\log(1-q)$ in the real case. Thus the log-determinant is biased downward by an amount proportional to $N$, while its fluctuations remain of order one. The bias is the Jensen gap of the logarithm integrated against the spreading of the Marchenko--Pastur bulk, and it is the seed of every information law in the next section. The exact formula \eqref{eq:logdet} and its limit underlie the high-dimensional entropy estimators of \citet{ahmed1989entropy}, \citet{misra2005estimation} and \citet{cai2015law}, and the corrected likelihood-ratio tests of \citet{bai2009corrections} and \citet{jiang2013central}.
\section*{Exact Spectral Laws of Information}

For a centred Gaussian vector with covariance $\Sigma$, the differential entropy is $h=\tfrac12\log\det(2\pi e\Sigma)$. The mutual information between the blocks of a partition $x=(x_1,x_2)$ of dimensions $(p,r)$ is
\[
 I(\Sigma)=\frac12\log\frac{\det\Sigma_{11}\det\Sigma_{22}}{\det\Sigma}
 =-\frac12\sum_{k=1}^{p}\log(1-\rho_k^2),
\]
where $\rho_k$ are the population canonical correlations \citep{cover2006elements}. The Kullback--Leibler divergence between two centred Gaussians is $\KL(\N_A\|\N_B)=\tfrac12[\tr(B^{-1}A)-N-\log\det(B^{-1}A)]$. All three are smooth functions of the covariance matrix and are routinely estimated by substituting $S$ for $\Sigma$. I now derive the exact and asymptotic laws of these plug-in estimators. Proofs that are not given in the text are in Appendix~A.

\subsection*{Spurious mutual information}

Define the digamma partial sums $\Psi_n(k)=\sum_{i=1}^k\psi((n-i+1)/2)$ and the \emph{spurious information}
\begin{equation}
 B_n(p,r)=\tfrac12\bigl[\Psi_n(p)+\Psi_n(r)-\Psi_n(p+r)\bigr]
 =\tfrac12\sum_{i=1}^{r}\Bigl[\psi\Bigl(\frac{n-i+1}2\Bigr)-\psi\Bigl(\frac{n-p-i+1}2\Bigr)\Bigr].
 \label{eq:spurious}
\end{equation}
The second expression follows from the first by telescoping, and it is symmetric in $(p,r)$ although it does not look it.

\begin{proposition*}[Exact spurious information]
Let $n\geq p+r$ and let $\hat I=I(S)$ be the plug-in mutual information computed from $S=W/n$ with $W\sim\mathcal W_{p+r}(n,\Sigma)$. For every positive definite $\Sigma$,
\[
 \E\hat I=I(\Sigma)+B_n(p,r).
\]
The excess $B_n(p,r)$ is strictly positive whenever $p,r\geq1$, and it does not depend on $\Sigma$. The bias-corrected estimator $\tilde I=\hat I-B_n(p,r)$ is exactly unbiased for $I(\Sigma)$.
\end{proposition*}

The proof is two lines. The diagonal blocks satisfy $nS_{11}\sim\mathcal W_p(n,\Sigma_{11})$ and $nS_{22}\sim\mathcal W_r(n,\Sigma_{22})$ with the same number of degrees of freedom \citep[Ch.~3]{muirhead1982aspects}. Applying \eqref{eq:logdet} to each of the three determinants in $\hat I$ gives $\E\hat I-I=\tfrac12[\ell_n(p)+\ell_n(r)-\ell_n(p+r)]$, and the $\log(n/2)$ terms cancel because $p+r-(p+r)=0$. Positivity follows from the second form of \eqref{eq:spurious} and the strict monotonicity of $\psi$ on $(0,\infty)$. The Lean file proves the telescoping identity and the positivity for an arbitrary strictly antitone sequence. The analytic content is confined to \eqref{eq:logdet} and to the monotonicity of the digamma function. The independence of $B_n$ from $\Sigma$ is what makes the correction feasible. An analyst does not need to know the dependence in order to remove the bias that dimensionality adds to its estimate. Nothing in this argument requires $p$ and $r$ to be small. The same proposition holds after demeaning with $n-1$ in place of $n$, and after regressing both blocks on $k$ common regressors with $n-k$ in place of $n$. The second case is the version used for partial mutual information in the application.

When $\Sigma_{12}=0$ the full null law is available. In that case $\Lambda=\det S/(\det S_{11}\det S_{22})=\prod_k(1-\hat\rho_k^2)$ is Wilks' statistic, and $\hat I=-\tfrac12\log\Lambda$.

\begin{proposition*}[Null law]
If $\Sigma_{12}=0$ and $n\geq p+r$, then $\hat I\stackrel{d}{=}-\tfrac12\sum_{i=1}^p\log\beta_i$ with independent $\beta_i\sim\Beta\bigl(\tfrac{n-r-i+1}2,\tfrac r2\bigr)$. Consequently
\[
 \E_0\hat I=B_n(p,r),\qquad
 \Var_0\hat I=V_n(p,r)\equiv\frac14\sum_{i=1}^p\Bigl[\psi'\Bigl(\frac{n-r-i+1}2\Bigr)-\psi'\Bigl(\frac{n-i+1}2\Bigr)\Bigr].
\]
\end{proposition*}

The product representation of Wilks' $\Lambda$ is classical \citep[Ch.~8]{anderson2003introduction}, and the moments follow from $\E\log\beta=\psi(a)-\psi(a+b)$ and $\Var\log\beta=\psi'(a)-\psi'(a+b)$ for $\beta\sim\Beta(a,b)$. The two expressions for $\E_0\hat I$ agree by the telescoping in \eqref{eq:spurious}, which provides an internal consistency check. The null law gives an exact one-sided test of informational independence, rejecting when $(\hat I-B_n)/V_n^{1/2}$ exceeds a normal or exact quantile. It also provides an exact standard error for the corrected estimator under the null.

\subsection*{Proportional asymptotics and the Wachter law}

The finite-sample formulas have simple limits when both blocks grow with the sample. Let
\begin{align}
 \kappa(c_1,c_2)&=\tfrac12\bigl[(1-c_1-c_2)\log(1-c_1-c_2)-(1-c_1)\log(1-c_1)-(1-c_2)\log(1-c_2)\bigr],
 \label{eq:kappa}\\
 \nu(c_1,c_2)&=\tfrac12\log\frac{(1-c_1)(1-c_2)}{1-c_1-c_2}.
 \label{eq:nu}
\end{align}

\begin{proposition*}[Proportional limits]
Let $p/n\to c_1$ and $r/n\to c_2$ with $c_1,c_2>0$ and $c_1+c_2<1$. Then $B_n(p,r)/n\to\kappa(c_1,c_2)$ and $V_n(p,r)\to\nu(c_1,c_2)$. Under independence, $(\hat I-B_n(p,r))/V_n(p,r)^{1/2}$ converges in distribution to a standard normal. For every $\Sigma$, $\hat I/n-I(\Sigma)/n-\kappa(c_1,c_2)\to0$ in probability.
\end{proposition*}

\begin{figure}[htbp]
\centering
\includegraphics[width=\textwidth]{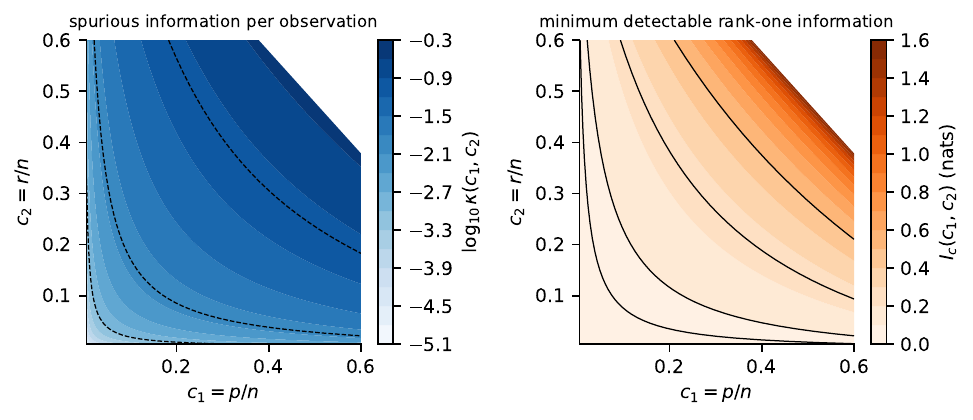}
\caption{Left: spurious information per observation, $\kappa(c_1,c_2)$ from \eqref{eq:kappa}, on a logarithmic colour scale, with contours at $10^{-3}$, $10^{-2}$ and $10^{-1}$. Right: the minimum detectable rank-one information $I_c(c_1,c_2)$ from \eqref{eq:frontier}, with contours at $0.05$, $0.1$, $0.25$ and $0.5$ nats. Both are defined on $c_1+c_2<1$.}
\label{fig:maps}
\end{figure}

The expansion $(1-c)\log(1-c)=-c+\sum_{k\ge2}c^k/[k(k-1)]$ gives the series
\begin{align*}
 \kappa(c_1,c_2)&=\frac12\sum_{k\geq2}\frac{(c_1+c_2)^k-c_1^k-c_2^k}{k(k-1)}
 =\frac{c_1c_2}2\Bigl(1+\frac{c_1+c_2}2+\cdots\Bigr),\\
 \nu(c_1,c_2)&=\frac12\sum_{k\ge2}\frac{(c_1+c_2)^k-c_1^k-c_2^k}{k}
 =\frac{c_1c_2}2\bigl(1+c_1+c_2+\cdots\bigr).
\end{align*}
All coefficients are positive, so $\kappa$ and $\nu$ are positive and increasing in each argument. The leading terms reproduce the classical fixed-dimension approximation $2n\hat I\approx\chi^2_{pr}$, with mean $pr/(2n)$ and variance $pr/(2n^2)$. The higher-order terms are the random matrix correction. They are not small. At $c_1=c_2=0.4$, the ratio $n\kappa/(pr/2n)$ equals $1.82$, so the classical formula misses almost half of the spurious information. Figure~\ref{fig:maps} maps $\kappa$ over the admissible region. The contours are hyperbola-like near the origin, where $\kappa\approx c_1c_2/2$, and steepen sharply as $c_1+c_2\to1$, where $(1-c_1-c_2)\log(1-c_1-c_2)$ has an infinite derivative.

The plug-in information is also a linear spectral statistic of the canonical correlations, $\hat I=-\tfrac p2\int\log(1-x)\,\hat F_p(\dd x)$, with $\hat F_p$ the empirical distribution of $\hat\rho_1^2,\ldots,\hat\rho_p^2$. Under independence, $\hat F_p$ converges weakly to the Wachter law \eqref{eq:wachter}, whose support stays inside $[0,\omega_+]$ with $\omega_+<1$. Combining this with the previous proposition identifies a logarithmic moment of the Wachter law that does not appear to have been recorded before.

\begin{theorem*}[Logarithmic moment of the Wachter law]
For $0<c_1\leq c_2$ with $c_1+c_2<1$,
\begin{multline*}
 \int_{\omega_-}^{\omega_+}\log(1-x)\,w_{c_1,c_2}(x)\,\dd x=-\frac2{c_1}\,\kappa(c_1,c_2)\\
 =\frac1{c_1}\Bigl[(1-c_1)\log(1-c_1)+(1-c_2)\log(1-c_2)-(1-c_1-c_2)\log(1-c_1-c_2)\Bigr].
\end{multline*}
\end{theorem*}

The proof in Appendix~A computes the same almost sure limit of $\hat I/n$ in two ways, once through the Bartlett decomposition and once through the Wachter law. It needs only the convergence of the extreme canonical correlations to the edges of the support. Numerical quadrature confirms the identity to $10^{-13}$ over a grid of $(c_1,c_2)$, and the comparison in Figure~\ref{fig:wachter} shows how quickly the empirical spectrum settles on the Wachter law. The same reasoning, applied to the variance $V_n$, identifies $\nu(c_1,c_2)$ as the limiting variance of the linear statistic $\sum_k\log(1-\hat\rho_k^2)/2$, in line with the central limit theorems for Jacobi ensembles used by \citet{jiang2013central} and \citet{yang2015independence}.

\subsection*{Uninformative coordinates}

An important practical consequence is that spurious information can be manufactured. Suppose an analyst adds $m$ coordinates to the second block that are independent of everything else, so that the true mutual information is unchanged.

\begin{proposition*}[Monotonicity in uninformative coordinates]
For $m\geq1$, $n\ge p+r+m$ and every $\Sigma$,
\begin{multline*}
 \E\hat I_{p,r+m}-\E\hat I_{p,r}=B_n(p,r+m)-B_n(p,r)\\
 =\frac12\sum_{i=1}^{m}\Bigl[\psi\Bigl(\frac{n-r-i+1}2\Bigr)-\psi\Bigl(\frac{n-p-r-i+1}2\Bigr)\Bigr]>0.
\end{multline*}
In the proportional regime with $m/n\to c_3$, the increment per observation tends to $\kappa(c_1,c_2+c_3)-\kappa(c_1,c_2)$.
\end{proposition*}

The identity is again a telescoping sum, proved in the Lean file for an arbitrary sequence, together with strict positivity for a strictly antitone one. Its economic content is developed in the next section. The point to note here is that the digamma-corrected estimator is flat in $m$ by construction, whereas regularised plug-in estimators are not. Shrinking the covariance matrix towards a multiple of the identity, as \citet{ledoit2004well} do, or applying the rotationally invariant estimator \eqref{eq:rie}, removes spurious information. It also removes genuine information, and the numerical section shows that their estimates of $I$ collapse towards zero as $m$ grows. Covariance cleaning is optimised for quadratic losses on $\Sigma$, and information is not a quadratic functional.

\subsection*{The information cost of plug-in beliefs}

A Gaussian decision maker who adopts the plug-in law $\N(0,S)$ in place of $\N(0,\Sigma)$ incurs two natural information costs. The forward divergence $\KL(\N_\Sigma\|\N_S)$ is the expected excess log loss, or code length in nats, per future observation of using the estimated density for prediction. The reverse divergence $\KL(\N_S\|\N_\Sigma)$ is one half of Stein's loss of $S$ \citep{james1961estimation}. Write $\ell_n(N)$ for the exact log-determinant bias \eqref{eq:logdet}.

\begin{proposition*}[Exact information costs]
For $n\geq N+2$ and every $\Sigma$,
\begin{align*}
 \E\,\KL(\N_\Sigma\|\N_S)&=\frac12\Bigl[\frac{N(N+1)}{n-N-1}+\ell_n(N)\Bigr],&
 \E\,\KL(\N_S\|\N_\Sigma)&=-\frac12\ell_n(N).
\end{align*}
As $N/n\to q\in(0,1)$, the costs per dimension converge to $\tfrac12[q/(1-q)+\varphi(q)]$ and $-\tfrac12\varphi(q)$ respectively, with $\varphi$ from \eqref{eq:phi}.
\end{proposition*}

The forward formula uses $\E W^{-1}=\Sigma^{-1}/(n-N-1)$ \citep[Ch.~3]{muirhead1982aspects} and the algebraic identity $Nn/(n-N-1)-N=N(N+1)/(n-N-1)$, which is certified in Lean. Both costs expand as $q/4+O(q^2)$ per dimension, so that for small $q$ the total cost is close to the parameter count $N(N+1)/2$ divided by $2n$, the familiar Akaike penalty. For larger $q$ the forward cost explodes like $(1-q)^{-1}$, because the predictive density inherits the smallest sample eigenvalues and those approach zero at the left Marchenko--Pastur edge. The reverse cost stays bounded as $q\to1$, with limit $\tfrac12$ per dimension. For the empirical design used later, with $N=49$ industries and $n=249$ demeaned daily observations, the exact forward cost is \KLfwdN{} nats and the reverse cost \KLrevN{} nats. With a sixty-day window ($n=59$), they rise to \KLfwdSixty{} and \KLrevSixty{} nats. Expressed per future day, the forward costs are large compared with the daily information content of most signals that finance tries to measure. The forward cost is the information-theoretic counterpart of the out-of-sample penalty for minimum-variance portfolios. For $w=S^{-1}\one/\one^\top S^{-1}\one$ the exact ratios of in-sample and out-of-sample variance to the true minimum variance have expectations
\begin{equation}
 \E\frac{R_{\rm in}}{R}=\frac{n-N+1}n,\qquad
 \E\frac{R_{\rm out}}{R}=\frac{n-1}{n-N},
 \label{eq:gmv}
\end{equation}
whose product tends to $(1-q)^{-2}$ \citep{kan2007optimal,elkaroui2010high,bodnar2018estimation}. The first identity follows from $\one^\top\Sigma^{-1}\one/\one^\top W^{-1}\one\sim\chi^2_{n-N+1}$ \citep[Ch.~3]{muirhead1982aspects}, and both are confirmed numerically below.

\subsection*{A frontier for detectable information}

The proportional limits describe the bulk. The spiked theory \eqref{eq:cca-spike} describes what can be learned about a single dependence direction. Suppose the cross-block dependence has rank one, with canonical correlation $t$, so that $I(\Sigma)=-\tfrac12\log(1-t^2)$. The leading sample canonical correlation carries information about $t$ only if $t>t_c$. Define the \emph{information frontier}
\begin{equation}
 I_c(c_1,c_2)=-\frac12\log\bigl(1-t_c^2\bigr)=-\frac12\log\Bigl(1-\sqrt{\frac{c_1c_2}{(1-c_1)(1-c_2)}}\Bigr).
 \label{eq:frontier}
\end{equation}
Rank-one information below $I_c$ is real, in the sense that it changes the joint law and has positive economic value, but it is invisible to the spectrum. The leading canonical correlation converges to $\omega_+$ whatever the value of $t$. The aggregate statistic $\tilde I$ still responds, since its mean shifts by $I(\Sigma)$ while its null standard deviation is $\nu^{1/2}$, but it does not by itself locate the information. Locating it is the task of the leading canonical pair, which fails below the threshold. The right panel of Figure~\ref{fig:maps} shows that $I_c$ is negligible when both blocks are small relative to the sample, and becomes substantial when either block is large. For example, $I_c=0.091$ nats at $(c_1,c_2)=(0.1,0.2)$, and $I_c$ exceeds half a nat when $c_1+c_2$ approaches $0.8$ with balanced blocks. The frontier gives a precise meaning to the intuition that, in large cross-sections, some information is real but cannot be acted upon.
\section*{The Value of Information and Its Spurious Component}

The information laws acquire economic meaning through a simple but exact identity. Consider a price-taking investor with constant absolute risk aversion $\gamma>0$ and initial wealth $w_0$ who can trade $N$ risky assets at prices $p\in\R^N$ and a riskless asset with zero net return. Terminal payoffs are $v\sim\N(\mu,\Sigma_v)$. Before trading, the investor may observe a signal $s\in\R^K$ that is jointly Gaussian with $v$. Let $\Sigma_{v|s}$ denote the conditional covariance of $v$ given $s$, which is non-random, and recall that $I(v;s)=\tfrac12\log\det(\Sigma_v\Sigma_{v|s}^{-1})$. Terminal wealth for a position $x$ is $w_0+x^\top(v-p)$.

\begin{theorem*}[Exact value of information]
Let $U_s$ and $U_\varnothing$ be the ex-ante expected utilities, under $u(w)=-e^{-\gamma w}$, of the optimal policies with and without the signal. Then
\[
 \frac{U_s}{U_\varnothing}=\exp\{-I(v;s)\},
\]
so that the certainty-equivalent value of the signal, the largest sum the investor would pay for it before observing it, is exactly $I(v;s)/\gamma$. The value is independent of $\mu$, $p$ and $w_0$, and is invariant under invertible linear transformations of $s$.
\end{theorem*}

Conditional on $s$, the optimal demand is $x^*=(\gamma\Sigma_{v|s})^{-1}(m-p)$ with $m=\E[v\mid s]$, and the conditional expected utility is $-\exp\{-\gamma w_0-\tfrac12(m-p)^\top\Sigma_{v|s}^{-1}(m-p)\}$. Ex ante, $m-p\sim\N(\mu-p,\Sigma_v-\Sigma_{v|s})$. For $y\sim\N(c,B)$ and positive semidefinite $A$, the Gaussian integral $\E\exp(-\tfrac12y^\top Ay)=\det(I+BA)^{-1/2}\exp\{-\tfrac12c^\top(A^{-1}+B)^{-1}c\}$ holds. With $A=\Sigma_{v|s}^{-1}$ and $B=\Sigma_v-\Sigma_{v|s}$ we have $I+BA=\Sigma_v\Sigma_{v|s}^{-1}$ and $A^{-1}+B=\Sigma_v$. The exponential factor is therefore exactly the uninformed expected utility, and the determinant factor is $e^{-I(v;s)}$. The scalar versions of the two matrix identities are certified in Lean, and Appendix~A gives the details.

The identity says that in the canonical model of portfolio choice under information, mutual information is not merely an analogy for value. It \emph{is} value, measured in units of risk tolerance. It makes precise the sense in which entropy-based measures order signals for investors \citep{cabrales2013entropy}, and it underlies the log-determinant objectives in models of information choice \citep{vannieuwerburgh2010information,veldkamp2011information,kacperczyk2016rational}. It also explains why a linear cost per nat, as in rational inattention \citep{sims2003implications,mackowiak2023rational}, produces bang-bang acquisition decisions in CARA--Gaussian settings: a signal is worth buying at a price of $\lambda$ per nat exactly when $1/\gamma>\lambda$. A further analogy is with \citet{kelly1956new}, who showed that a log-utility gambler with side information grows wealth faster by exactly the mutual information. The theorem is the CARA counterpart, with certainty equivalents in place of growth rates. The identity is partial-equilibrium. In a noisy rational expectations equilibrium prices themselves carry information \citep{grossman1980impossibility,hellwig1980aggregation,verrecchia1982information}, and the relevant quantity becomes the conditional mutual information $I(v;s\mid p)$. The theorem applies verbatim to that quantity because conditioning on a jointly Gaussian price only replaces $\Sigma_v$ by $\Sigma_{v|p}$.

\subsection*{Spurious value of information}

Valuing a signal in practice requires estimating $I(v;s)$. An investor, an econometrician or a data vendor with $n$ joint observations of $(v,s)$, with $\dim v=N$ and $\dim s=K$, who uses the plug-in estimate reports $\hat V=\hat I/\gamma$. All results of the previous section transfer. The expected excess of the reported value over the true one is
\[
 \E\hat V-V=\frac{B_n(N,K)}\gamma>0,
\]
exactly, and the corrected value $\tilde V=(\hat I-B_n(N,K))/\gamma$ is unbiased. Under the null of a worthless signal, $\gamma\hat V$ has the log-Beta law of the null proposition, so a test for positive value is available at any sample size. In the proportional regime the spurious value grows like $n\kappa(N/n,K/n)/\gamma$. It is therefore not a small-sample nuisance that vanishes with more data. If the numbers of assets and signals grow with the sample, it is a permanent wedge, whose size per observation is set by $\kappa$ and shown in Figure~\ref{fig:maps}. With a risk tolerance of one million dollars, the spurious value in the empirical design of the application, with eight assets, four signals and a year of daily data, is $\$\,$\MIBias{}~million, or about $66{,}000$ dollars per decision.

The monotonicity proposition has a sharper implication for markets in which information is sold. Suppose a vendor offers $K$ signals and a buyer evaluates them by the plug-in value on a common sample. Appending $m$ series that are independent of the payoffs and of the original signals leaves the true value unchanged. It raises the expected measured value by
\begin{multline*}
 \Delta_m=\frac{B_n(N,K+m)-B_n(N,K)}\gamma\\
 =\frac1{2\gamma}\sum_{i=1}^m\Bigl[\psi\Bigl(\frac{n-K-i+1}2\Bigr)-\psi\Bigl(\frac{n-N-K-i+1}2\Bigr)\Bigr]>0.
\end{multline*}
A buyer who does not correct for dimension therefore faces a lemons problem in the sense of \citet{akerlof1970market}, in which padding is privately profitable and undetectable on average. The correction restores the right incentives in expectation. It cannot restore precision. The null variance $V_n(N,K+m)$ is strictly increasing in $m$, because $\psi'$ is decreasing, so padded signal sets remain harder to evaluate even when evaluated without bias. This mechanism offers a formal account of a concern that runs through work on data markets and financial data technology \citep{farboodi2020long}: as the number of available series grows, the measured information content of data rises mechanically, while the economically relevant content need not. It is also a precise sense in which uninformative or polluted information sources degrade price discovery. They do not mislead on average once the analyst corrects, but they consume degrees of freedom and inflate uncorrected assessments.

\subsection*{Information that cannot be located}

The frontier \eqref{eq:frontier} adds a further layer. Suppose the dependence between payoffs and signals is concentrated in one direction, so that the true value is $-\tfrac1{2\gamma}\log(1-t^2)$. An investor who knows the direction can trade on it and capture that value. A statistician who must learn the direction from $n$ observations of $N$ payoffs and $K$ signals faces the canonical-correlation phase transition. When $t<t_c$, the leading sample canonical correlation converges to the bulk edge $\omega_+$ whatever the value of $t$, so it reveals nothing about $t$. By analogy with the eigenvector part of the BBP transition, the estimated canonical direction is then expected to carry no asymptotic information about the true direction either. I verify the eigenvalue statement by simulation below, and leave the eigenvector statement as a conjecture supported by the analogy. The value $V_c=I_c(N/n,K/n)/\gamma$ is thus the largest rank-one value that remains inaccessible to learning from data at the given dimensions. Advantages below that level can persist without being competed away by statistical arbitrage. This gives an information-theoretic counterpart to the observation of \citet{martin2022market} that investors who learn in high dimensions face estimation frictions that make returns appear predictable ex post. It also connects to the evidence of \citet{kelly2024virtue} and \citet{kozak2020shrinking} that the relevant signals in asset pricing are numerous and individually weak. In the empirical design used below, the frontier is small, $I_c=\FrontierEmp{}$ nats with $t_c=\FrontierEmpT{}$, because eight and four are small compared with a year of daily data. At the scale of the full cross-section of individual stocks, with hundreds of characteristics, it becomes first order.
\section*{Spectral Dynamics and Information Rates}

Correlation matrices of financial returns fluctuate, and much of econophysics concerns the dynamics of their spectra. Examples include the motion of the market mode, the reshuffling of sector eigenvectors and the alternation of correlation regimes \citep{allez2012eigenvector,munnix2012identifying,schmitt2013non,reigneron2011principal}. The simplest model that has the right symmetries is a symmetric matrix Ornstein--Uhlenbeck process,
\begin{equation}
 \dd H_t=-a\,H_t\,\dd t+\frac1{\sqrt N}\,\dd B_t,
 \label{eq:mou}
\end{equation}
where $B$ is a symmetric matrix Brownian motion whose entries $B_{ij}$, $i<j$, are independent standard Brownian motions and whose diagonal entries are $\sqrt2$ times independent standard Brownian motions. I interpret $H_t$ as the deviation of an $N\times N$ correlation or covariance matrix from its long-run level, with $a>0$ the speed at which correlation shocks dissipate. Matrix-valued diffusions of this kind and their Wishart counterparts have been studied by \citet{bru1991wishart} and \citet{gourieroux2009wishart}. Their eigenvalue dynamics go back to \citet{dyson1962brownian}, and the convergence of their spectra to the semicircle law to \citet{chan1992wigner} and \citet{rogers1993interacting}. The stationary law of \eqref{eq:mou} is the Gaussian orthogonal ensemble scaled so that off-diagonal entries have variance $1/(2aN)$ and diagonal entries $1/(aN)$.

\subsection*{Dyson dynamics and the Burgers equation}

\begin{figure}[htbp]
\centering
\includegraphics[width=\textwidth]{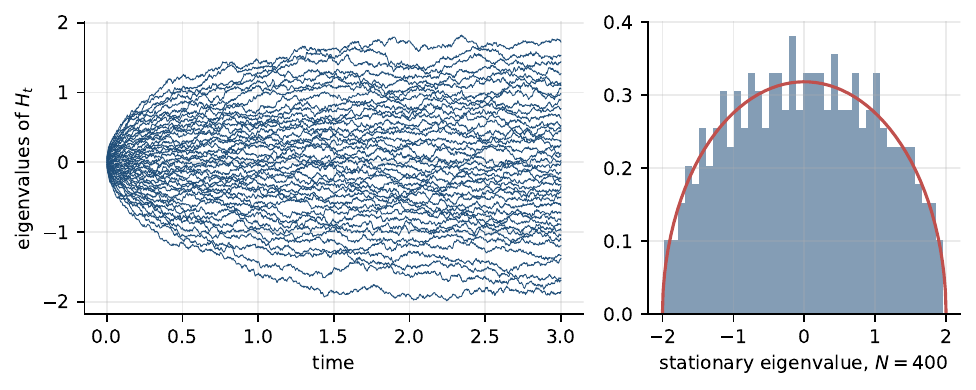}
\caption{Left: eigenvalue trajectories of the matrix Ornstein--Uhlenbeck process \eqref{eq:mou} with $a=\tfrac12$ and $N=40$, started from $H_0=0$ and simulated by Euler steps of length $0.002$. The trajectories repel and never cross. Right: stationary spectrum of a $400\times400$ draw against the semicircle of radius $\sqrt{2/a}=2$, the stationary solution of the Burgers equation \eqref{eq:burgers}.}
\label{fig:dyson}
\end{figure}

The eigenvalues $\lambda_1(t)>\cdots>\lambda_N(t)$ of $H_t$ satisfy the Dyson system
\begin{equation}
 \dd\lambda_i=\sqrt{\frac2N}\,\dd w_i+\Bigl[\frac1N\sum_{j\neq i}\frac1{\lambda_i-\lambda_j}-a\lambda_i\Bigr]\dd t,
 \label{eq:dyson}
\end{equation}
with independent Brownian motions $w_i$. For the orthogonal symmetry class the eigenvalues never collide after time zero \citep[Section 4.3]{anderson2010introduction}, so \eqref{eq:dyson} has a unique strong solution. The repulsion term is the second-order perturbation of each eigenvalue by the others. The confinement term is the image of the matrix drift, since $-aH$ commutes with every rotation and therefore acts on the spectrum alone. Let $g_t(z)=N^{-1}\tr(z-H_t)^{-1}$. Applying It\^o's formula to $(z-\lambda_i)^{-1}$, symmetrising the repulsion sum through $\sum_{i\ne j}f'(\lambda_i)/(\lambda_i-\lambda_j)=\tfrac12\sum_{i\neq j}[f'(\lambda_i)-f'(\lambda_j)]/(\lambda_i-\lambda_j)$ and noting that the It\^o correction is $O(N^{-1})$, gives in the large-$N$ limit the forced complex Burgers equation
\begin{equation}
 \partial_tg=-g\,\partial_zg+a\,\partial_z(zg).
 \label{eq:burgers}
\end{equation}
The first term is the inviscid Burgers flow of free Brownian motion, whose characteristics are straight lines. In terms of the $R$-transform, it states that the free cumulants of the spectrum increase linearly in time, $R_t(w)=R_0(w)+tw$, which is the free-probability analogue of variance adding under independent increments. The second term is the deterministic contraction. A stationary solution must satisfy $g\,\partial_zg=a\,\partial_z(zg)$. Integrating once and using $g\sim1/z$ at infinity gives $g^2=2a(zg-1)$, so
\[
 g_\infty(z)=az-\sqrt{a^2z^2-2a},\qquad \rho_\infty(x)=\frac1\pi\sqrt{2a-a^2x^2},\quad|x|\le\sqrt{2/a},
\]
the semicircle law of radius $\sqrt{2/a}$. For $a=\tfrac12$ this is the standard semicircle on $[-2,2]$. Started from $H_0=0$, the solution of \eqref{eq:burgers} is at each time a semicircle with variance $\sigma_t^2=(1-e^{-2at})/(2a)$, which interpolates between free Brownian motion at small times and the stationary law. Figure~\ref{fig:dyson} shows a simulated path of the Dyson system for $N=40$ together with the stationary spectrum at $N=400$. The level repulsion visible in the left panel is the mechanism by which sample eigenvalues avoid crossing, and it is the dynamic counterpart of the Marchenko--Pastur spreading.

\subsection*{Information rates and sufficiency of the spectrum}

Suppose an observer monitors $H$ continuously over $[0,T]$ and must decide whether correlation shocks dissipate at rate $a$ or at rate $b$. Let $P_a^T$ and $P_b^T$ be the corresponding laws of the path $(H_t)_{t\le T}$, both started from the stationary law of the $a$-regime. Let $\Lambda_a^T$ and $\Lambda_b^T$ be the laws of the eigenvalue path $(\lambda(t))_{t\le T}$ alone.

\begin{theorem*}[Information rate and sufficiency of the Dyson process]
For every $N\ge1$ and $T>0$,
\[
 \KL(P_a^T\|P_b^T)=\KL(\Lambda_a^T\|\Lambda_b^T)=T\,\frac{N(N+1)(b-a)^2}{8a}.
\]
Consequently the eigenvector process carries no information about the dissipation rate. By the Chernoff--Stein lemma, the error exponent for discriminating the two regimes from continuous observation is $N(N+1)(b-a)^2/(8a)$ per unit time.
\end{theorem*}

The matrix part is Girsanov's theorem applied to the $N(N+1)/2$ independent coordinates of \eqref{eq:mou} \citep{girsanov1960transforming}. Each off-diagonal coordinate has diffusion variance $1/N$ per unit time and drift difference $(b-a)H_{ij}$. Each diagonal coordinate has diffusion variance $2/N$ and drift difference $(b-a)H_{ii}$. The Kullback--Leibler divergence is half the expected integrated squared drift difference, measured in the inverse diffusion metric:
\[
 \KL(P_a^T\|P_b^T)=\frac N4(b-a)^2\int_0^T\E_a\tr H_t^2\,\dd t,\qquad
 \E_a\tr H_t^2=\frac{N(N-1)}{2aN}+\frac N{aN}=\frac{N+1}{2a}.
\]
The eigenvalue part applies Girsanov's theorem to \eqref{eq:dyson}. The two regimes share the repulsion term and differ only in the confinement, with drift difference $(b-a)\lambda_i$ against diffusion variance $2/N$. This gives $\tfrac N4(b-a)^2\int\E_a\sum_i\lambda_i^2\,\dd t$, which is the same integral because $\sum_i\lambda_i^2=\tr H^2$. The two divergences therefore coincide. Since the eigenvalue path is a measurable function of the matrix path, the data-processing inequality says that equality of divergences is exactly the statement that no information about $a$ is lost by discarding the eigenvectors. The Lean file proves both closed forms and their equality as identities in the field of rationals. The Chernoff--Stein statement is the continuous-time version of the classical result that the optimal type-II error exponent at fixed type-I error equals the Kullback--Leibler rate \citep{chernoff1952measure,cover2006elements}.

Two consequences matter for empirical work. First, the time needed to distinguish two dissipation regimes at a given error level scales as $8a\log(1/\varepsilon)/[N(N+1)(b-a)^2]$. It falls quadratically with the number of assets, because each of the $N(N+1)/2$ independent correlation coordinates contributes its own evidence and rotation invariance concentrates all of that evidence in the $N$ eigenvalues. For $N=10$, $a=\tfrac12$ and $b=1$ the rate is \MOUrate{} nats per unit time, and simulation of the Euler log-likelihood ratio reproduces it in Table~\ref{tab:mc}. Second, the result is a sufficiency statement about a symmetry. If the drift breaks rotation invariance, for example $-a(H-M)$ with a fixed non-scalar target $M$, then the eigenvector process is informative about $M$. The rate for discriminating two targets $M$ and $M'$ at a common $a$ is $\tfrac{a^2N}4\|M-M'\|_F^2$, which includes rotations of $M$ that leave its spectrum unchanged and are therefore invisible to $\lambda(t)$.

The path-space divergence in the theorem is the same object that regularises variational inference for latent stochastic differential equations. There the evidence lower bound subtracts $\tfrac12\E\int\|u_t\|^2\dd t$, with $u$ the recognition drift expressed in the coordinates in which the diffusion is standard \citep{li2020scalable,rice2026sldi}. The sufficiency result therefore has a direct computational corollary for matrix-valued latent states. If both the generative and the recognition drifts are rotation-equivariant spectral functions of $H$, the Kullback--Leibler term and the likelihood of any rotation-invariant observation depend on the eigenvalue path alone. Variational inference can then be carried out on the $N$-dimensional Dyson system rather than on the $N(N+1)/2$ matrix coordinates, without loss. At a finite time step $\Delta$, the Euler transition laws under the two regimes share their Gaussian covariance and differ only in their means. The chain rule for relative entropy therefore gives a per-step divergence of $\tfrac N4(b-a)^2\Delta\tr H^2$ at the current state, both for the matrix chain and, with $\sum_i\lambda_i^2$ in place of $\tr H^2$, for the eigenvalue chain. The two discrete rates agree up to the $O(\Delta)$ discrepancy between the stationary second moments of the two Euler chains. The stationary initial laws contribute an additional, horizon-independent divergence of $\tfrac{N(N+1)}4[b/a-1-\log(b/a)]$ when the two regimes are started from their own stationary laws. Because this term grows like $N^2$ as well, cross-sectional information about correlation regimes is abundant both in snapshots and in dynamics.
\section*{Numerical Verification}

Every exact formula and every limit in the preceding sections was checked by Monte Carlo simulation with fixed seeds. Table~\ref{tab:mc} reports the comparison. Each row uses \MCReps{} independent Wishart draws, except the spiked rows (400 draws at $N=400$, $n=1600$) and the dynamic row (200 Euler paths). The covariance matrices in the rows that allow a general $\Sigma$ are random, with condition numbers of 10 or 20, so the absence of $\Sigma$ from the exact formulas is itself being tested. The last column reports the proportional limit where one exists, which shows how far finite-sample values sit from their limits at the dimensions used. The exact second and third Wishart moments agree to four decimals, while the Narayana limit $1+3q+q^2$ is off by about one hundredth at $n=400$. That gap is precisely the $3(1+q)/n+4/n^2$ correction. Measured against the limit instead of the exact value, the simulated third moment would appear to deviate by more than ten standard errors, a small reminder of the theme of this paper. The log-determinant row shows that the second-order term $\tfrac12\log(1-q)$ is needed to bridge the limit $N\varphi(q)$ and the exact value. Every other exact entry lies within two Monte Carlo standard errors of its simulated counterpart. This includes the spurious information under strong dependence, its null mean and Wilks variance, both Kullback--Leibler costs, both minimum-variance ratios, the BBP outlier and overlap at two spike strengths, and the matrix Ornstein--Uhlenbeck information rate.

\begin{table}[htbp]
\centering
\caption{Monte Carlo verification of the exact and limiting formulas. ``Exact or limit'' is the analytic value used for the comparison. For the spiked rows and the dynamic row this is the large-$N$ limit or the continuous-time value. The final column gives the proportional limit where it differs. Dimensions: moments and log-determinant at $N=100$, $n=400$. Information rows at $(p,r,n)=(10,20,100)$. Kullback--Leibler rows at $N=20$, $n=100$. Portfolio rows at $N=50$, $n=200$. BBP rows at $N=400$, $n=1600$. Matrix OU at $N=10$, $a=\tfrac12$, $b=1$, Euler step $0.01$ over $20$ time units.}
\label{tab:mc}
\small
\begin{tabular}{lrrrr}
\toprule
Quantity & Exact or limit & Simulated & MC s.e. & Limit \\
\midrule
\multicolumn{5}{l}{\emph{Spectral moments and log-determinant}}\\[1pt]
\quad $N^{-1}\E\tr S^2$ (exact) & 1.2525 & 1.2521 & 0.0003 & 1.2500 \\
\quad $N^{-1}\E\tr S^3$ (exact) & 1.8219 & 1.8212 & 0.0007 & 1.8125 \\
\quad $\E\log\det S-\log\det\Sigma$ (exact) & \ensuremath{-}13.8391 & \ensuremath{-}13.8515 & 0.0119 & \ensuremath{-}13.6954 \\
\midrule
\multicolumn{5}{l}{\emph{Spurious information}}\\[1pt]
\quad $\E\hat I-I$, $(p,r,n)=(10,20,100)$ & 1.1904 & 1.1885 & 0.0037 & 1.1833 \\
\quad $\E_0\hat I$ (independence) & 1.1904 & 1.1871 & 0.0019 & -- \\
\quad $\Var_0\hat I$ (Wilks) & 0.0143 & 0.0140 & 0.0003 & 0.0141 \\
\midrule
\multicolumn{5}{l}{\emph{Information costs and portfolio risk}}\\[1pt]
\quad $\E\KL(\N_\Sigma\|\N_S)$ & 1.5283 & 1.5320 & 0.0035 & 1.4257 \\
\quad $\E\KL(\N_S\|\N_\Sigma)$ & 1.1299 & 1.1312 & 0.0018 & 1.0743 \\
\quad $\E R_{\rm in}/R$ & 0.7550 & 0.7535 & 0.0014 & 0.7500 \\
\quad $\E R_{\rm out}/R$ & 1.3267 & 1.3279 & 0.0012 & 1.3333 \\
\midrule
\multicolumn{5}{l}{\emph{Spiked covariance (BBP)}}\\[1pt]
\quad BBP outlier, $\ell=2$ & 2.5000 & 2.5043 & 0.0031 & -- \\
\quad BBP overlap$^2$, $\ell=2$ & 0.6000 & 0.5985 & 0.0022 & -- \\
\quad BBP outlier, $\ell=4$ & 4.3333 & 4.3263 & 0.0071 & -- \\
\quad BBP overlap$^2$, $\ell=4$ & 0.8974 & 0.8974 & 0.0004 & -- \\
\midrule
\multicolumn{5}{l}{\emph{Matrix Ornstein--Uhlenbeck dynamics}}\\[1pt]
\quad matrix OU KL rate & 6.8750 & 6.9439 & 0.0873 & -- \\
\bottomrule
\end{tabular}
\end{table}

\begin{figure}[htbp]
\centering
\includegraphics[width=\textwidth]{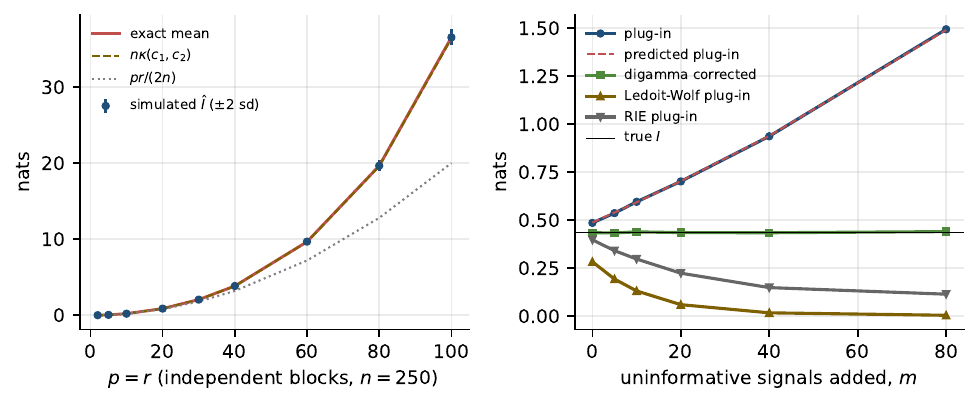}
\caption{Spurious information. Left: plug-in mutual information of independent blocks with $p=r$ and $n=250$, with the exact mean $B_n(p,p)$, the proportional limit $n\kappa(p/n,p/n)$ and the classical $pr/(2n)$. Right: a design with five payoffs, five informative signals with canonical correlation $0.4$, and $m$ appended uninformative signals, at $n=250$. The plug-in estimate rises as predicted by the monotonicity proposition, the digamma-corrected estimate stays at the true value, and the Ledoit--Wolf and RIE plug-in estimates are biased towards zero.}
\label{fig:spurious}
\end{figure}

The left panel of Figure~\ref{fig:spurious} plots the null mean of the plug-in information for independent blocks of equal size $p=r$ at $n=250$, with $\pm2$ standard deviation bands. The exact digamma formula and the proportional limit $n\kappa$ are indistinguishable at this scale and pass through the simulated means. The classical approximation $pr/(2n)$ is accurate for $p\leq20$ and then falls behind, reaching about $55\%$ of the true spurious information at $p=r=100$. The bands also illustrate the second message of the proportional limit. The standard deviation grows from $0.006$ at $p=2$ to only $0.54$ at $p=100$, so the spurious information becomes overwhelmingly significant as dimension grows. An uncorrected test would reject independence with probability approaching one. Table~\ref{tab:spurious} in Appendix~C reports the underlying numbers, including the exact and simulated standard deviations.

The right panel of Figure~\ref{fig:spurious} and Table~\ref{tab:noise} illustrate the monotonicity proposition and its economic reading. Five payoffs are linked to five informative signals with canonical correlations of $0.4$, so the true information is $0.436$ nats. Appending $m$ independent signals leaves the truth unchanged. The plug-in estimate nevertheless climbs from $0.485$ at $m=0$ to $1.494$ at $m=80$, matching the exact prediction $I+B_n(5,5+m)$ to within simulation error at every $m$. The digamma-corrected estimate stays between $0.433$ and $0.440$. The two cleaning estimators behave very differently. Ledoit--Wolf shrinkage removes the spurious information and much of the genuine information too, so its estimate falls to $0.004$ nats at $m=80$. The rotationally invariant estimator is less aggressive but also drifts downward, to $0.114$ nats. Shrinkage towards the identity pulls all canonical correlations towards zero, and adding noise coordinates raises the optimal shrinkage intensity. For the estimation of information, exact bias correction dominates covariance cleaning.

\begin{figure}[htbp]
\centering
\includegraphics[width=0.68\textwidth]{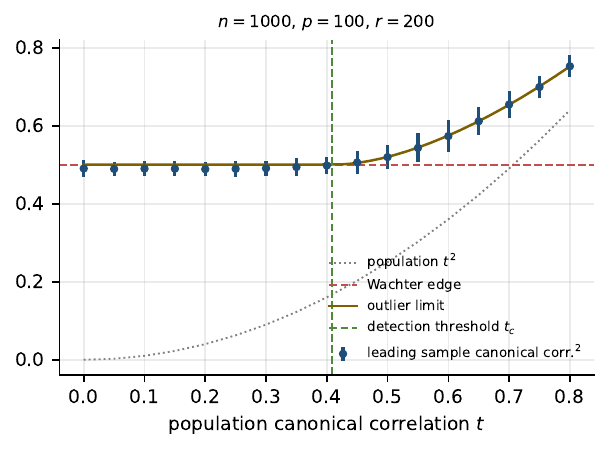}
\caption{Canonical-correlation phase transition at $n=1000$, $p=100$, $r=200$. The leading squared sample canonical correlation (mean $\pm2$ s.d. over 160 draws) sticks to the Wachter edge $\omega_+=\CCAEdge$ below the threshold $t_c=\CCAThr$ and follows the outlier limit $\gamma(t)$ of \eqref{eq:cca-spike} above it. Below the threshold, finite-$N$ draws sit slightly below $\omega_+$, as the Tracy--Widom scaling predicts.}
\label{fig:cca}
\end{figure}

Figure~\ref{fig:cca} checks the canonical-correlation transition behind the information frontier. With $c_1=0.1$ and $c_2=0.2$, the threshold is $t_c=\CCAThr$ and the Wachter edge is $\omega_+=\CCAEdge$. Below the threshold the leading squared sample canonical correlation stays at the edge, at about $0.49$ because of the negative finite-$N$ edge shift, whatever the population value. Above it, the simulated means follow $\gamma(t)$ closely. At $t=0.6$ the formula gives \CCAsixA{} and the simulation \CCAsixS{}, and at $t=0.8$ they are \CCAeightA{} and \CCAeightS{}. The information content of the threshold direction is $I_c=\FrontierSim{}$ nats. A rank-one link weaker than that is invisible to the leading canonical correlation at these dimensions, even though its value to an investor who knew its direction would be about nine percent of a unit of risk tolerance.
\section*{Application: Spectra and Information in US Industry Returns}

The application uses daily value-weighted returns of the \NAssets{} Fama--French industry portfolios from \SampleStart{} to \SampleEnd{}, a total of \NDays{} trading days, taken from the CRSP 202609 vintage of the Kenneth French data library \citep{fama1997industry,french2026data}. All 49 series are complete over this period. Industry portfolios are a natural laboratory for econophysics. They are numerous enough for random matrix effects to matter at common window lengths, the economic content of their eigenvectors is easy to read, and they are diversified enough that the extreme idiosyncratic events which dominate individual stocks do not drive the spectrum. Returns are heavy tailed and heteroskedastic \citep{cont2001empirical}, so the Gaussian identities of the previous sections are benchmarks here rather than literal descriptions. Part of the purpose of the application is to see how well they survive. All statistics are computed in rolling windows of $T=250$ trading days, advanced in steps of 21 days, which gives \NWindows{} windows with aspect ratio $q=49/250=0.196$ for the correlation analysis.

\subsection*{The spectrum of industry correlations}

\begin{figure}[htbp]
\centering
\includegraphics[width=0.8\textwidth]{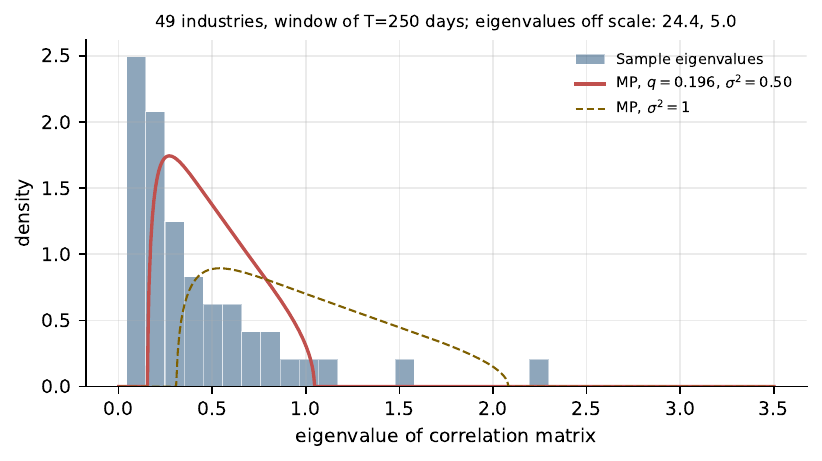}
\caption{Eigenvalues of the 49-industry correlation matrix over calendar year 2025 ($T=250$, $q=\SnapQ$), with the Marchenko--Pastur density rescaled by the residual variance $\hat\sigma^2=1-\lambda_1/N$ and the unscaled density for reference. The market mode ($\lambda_1=\SnapLambdaOne$) and the second eigenvalue lie off the scale of the plot.}
\label{fig:snapshot}
\end{figure}

Figure~\ref{fig:snapshot} shows the eigenvalues of the correlation matrix estimated over calendar year 2025. Following \citet{laloux1999noise}, the Marchenko--Pastur benchmark rescales the bulk variance to $\hat\sigma^2=1-\lambda_1/N=\SnapSigma$, the variance not absorbed by the market mode, which places the upper edge at \SnapEdge. The largest eigenvalue is \SnapLambdaOne{}, so the market mode carries half of the total variance. In total, \SnapNOut{} eigenvalues exceed the rescaled edge. The bulk is more concentrated near zero than the white Marchenko--Pastur law predicts. This is the signature of free multiplicative convolution with a non-flat population spectrum, $H\boxtimes\rho_q$ in the notation of the preliminaries. Industry portfolios have heterogeneous residual variances and some near-collinear pairs, so the population residual spectrum $H$ is itself spread out, and the sample bulk inherits that spread. The white benchmark is therefore conservative for detecting outliers at the upper edge and uninformative about the lower edge.

The BBP inversion turns each outlier into a statement about reliability. Table~\ref{tab:bbp} reports, for the eight largest eigenvalues of the 2025 window, the eigenvalue scaled by $\hat\sigma^2$, the population spike $\hat\ell_k$ implied by \eqref{eq:bbp}, and the predicted squared overlap \eqref{eq:overlap} between the sample eigenvector and its population counterpart. The market mode is estimated almost perfectly. The second mode, which loads most heavily on consumer staples (soft drinks, food, household products and beer), has a predicted overlap of $0.98$. The third and fourth modes, which isolate tobacco, entertainment and defence on the one hand and precious metals, coal and mining on the other, remain well identified with overlaps of $0.93$ and $0.84$. The fifth mode is a borderline case. Its implied spike of $1.85$ lies only modestly above the threshold $1+\sqrt q=1.44$, and its predicted overlap of $0.59$ means that about forty percent of the variance of the estimated direction is noise. Its loadings mix agriculture, insurance and beverages and should not be interpreted economically. The sixth and higher eigenvalues are inside the scaled edge, and the theory predicts that their eigenvectors are uninformative. This kind of triage is the practical value of the outlier theory. It replaces the binary question of whether an eigenvalue is above the edge with a graded measure of how much the eigenvector can be trusted.

\begin{table}[htbp]
\centering
\caption{Spiked-model triage of the eight largest eigenvalues of the 2025 industry correlation matrix ($q=\SnapQ$, $\hat\sigma^2=\SnapSigma$). Scaled eigenvalues are $\lambda_k/\hat\sigma^2$ for $k\geq2$. The implied spike solves $\ell^2-(\lambda+1-q)\ell+\lambda=0$, and the predicted overlap is \eqref{eq:overlap}. A dash means the scaled eigenvalue is inside the Marchenko--Pastur bulk. Loadings are the four industries with the largest absolute eigenvector weights.}
\label{tab:bbp}
\small
\begin{tabular}{rrrrrl}
\toprule
$k$ & $\lambda_k$ & $\lambda_k/\hat\sigma^2$ & implied $\ell_k$ & predicted $|\langle u_k,v_k\rangle|^2$ & largest loadings\\
\midrule
1 & 24.38 & 24.38 & 24.18 & 0.99 & BldMt, Chems, Fin, Banks \\
2 & 5.02 & 9.98 & 9.76 & 0.98 & Soda, Food, Hshld, Beer \\
3 & 2.30 & 4.57 & 4.32 & 0.93 & Smoke, Fun, Guns, Txtls \\
4 & 1.55 & 3.08 & 2.77 & 0.84 & Gold, Coal, Mines, Rtail \\
5 & 1.14 & 2.28 & 1.85 & 0.59 & Agric, Insur, Beer, Soda \\
6 & 1.04 & 2.07 & -- & 0.00 & Guns, Coal, Oil, Drugs \\
7 & 0.90 & 1.80 & -- & 0.00 & Gold, Oil, Fun, Aero \\
8 & 0.85 & 1.69 & -- & 0.00 & Guns, Drugs, Oil, MedEq \\
\bottomrule
\end{tabular}
\end{table}

\begin{figure}[htbp]
\centering
\includegraphics[width=\textwidth]{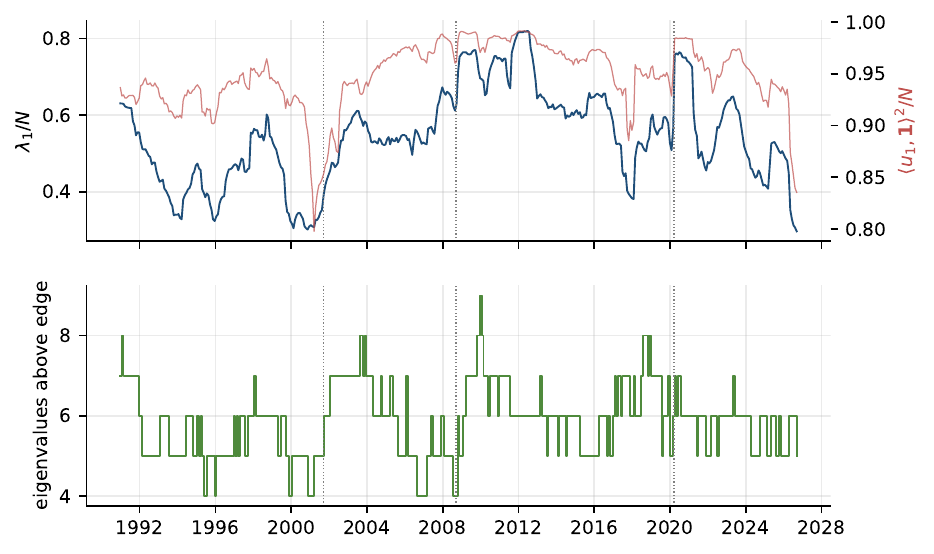}
\caption{Rolling spectral statistics, 250-day windows advanced by 21 days. Top: absorption ratio $\lambda_1/N$ (left axis) and the squared overlap of the market eigenvector with the normalised equal-weighted vector (right axis). Bottom: number of eigenvalues above the rescaled Marchenko--Pastur edge. Dotted lines mark 11 September 2001, 15 September 2008 and 16 March 2020.}
\label{fig:rolling}
\end{figure}

Over the full sample the picture is stable in structure and variable in intensity. Figure~\ref{fig:rolling} shows the absorption ratio $\lambda_1/N$, which is the share of variance carried by the market mode and the systemic-risk indicator of \citet{kritzman2011principal}, together with the squared overlap between the market eigenvector and the equal-weighted direction and the number of eigenvalues above the edge. Absorption averages \AbsMean{} and ranges from \AbsMin{} in the window ending in \AbsMinDate{} to \AbsMax{} in the window ending in \AbsMaxDate{}. It rises sharply around the market dislocations of 2001, 2008 and 2020 and stays elevated through the euro-area crisis. The market eigenvector never strays far from the equal-weighted direction, with a squared overlap of at least \OverlapMin{}. Its overlap tends to fall when absorption falls, as the BBP overlap formula predicts for a weaker spike. The number of outliers varies between \OutMin{} and \OutMax{}, with a median of \OutMedian{}, and is positively associated with absorption (correlation \CorrAbsOut). Crises strengthen the market mode and also separate additional sector modes from the bulk. This is consistent with the regime picture of \citet{munnix2012identifying} and with the dynamic analysis above, in which correlation shocks are spectral events that move several eigenvalues together.

\subsection*{The out-of-sample cost of noise dressing}

The minimum-variance portfolio is the cleanest setting in which the cost of estimation noise can be read off directly, because it depends on the covariance matrix alone. I form global minimum-variance portfolios of the 49 industries every 21 trading days from estimation windows of $T\in\{60,120,250,500\}$ days and hold them for the following 21 days. Every estimator is evaluated on the same out-of-sample period, from the 500th trading day of the sample onward. The estimators are the sample covariance matrix, eigenvalue clipping of the correlation matrix at the rescaled Marchenko--Pastur edge \citep{laloux2000random}, linear shrinkage towards a scaled identity \citep{ledoit2004well}, and the rotationally invariant estimator \eqref{eq:rie} of \citet{bun2016rotational} with $\eta=N^{-1/2}$. Table~\ref{tab:backtest} reports realised annualised volatilities. It also reports, for the sample estimator, the ratio of realised to predicted portfolio variance next to the Gaussian benchmark $(1-q)^{-2}$ implied by \eqref{eq:gmv}.

\begin{table}[htbp]
\centering
\caption{Global minimum-variance portfolios of 49 industries, rebalanced every 21 days on a common out-of-sample period. Columns 3--6: realised annualised volatility in percent. Column 7: ratio of realised to in-sample predicted variance for the sample estimator. Column 8: its Gaussian benchmark $(1-q)^{-2}$, the product of the in-sample and out-of-sample factors in \eqref{eq:gmv}.}
\label{tab:backtest}
\small
\begin{tabular}{rrrrrrrr}
\toprule
$T$ & $q$ & Sample & Clipped & Ledoit--Wolf & RIE & $\hat R_{\rm out}/\hat R_{\rm in}$ & $(1-q)^{-2}$\\
\midrule
60 & 0.817 & 23.51 & 11.87 & 11.99 & 15.37 & 44.07 & 29.75 \\
120 & 0.408 & 13.02 & 11.84 & 11.59 & 12.06 & 3.77 & 2.86 \\
250 & 0.196 & 11.76 & 11.72 & 11.36 & 11.39 & 2.06 & 1.55 \\
500 & 0.098 & 11.59 & 11.93 & 11.46 & 11.47 & 1.63 & 1.23 \\
\midrule
\multicolumn{8}{l}{Equal weight ($1/N$): 17.93}\\
\bottomrule
\end{tabular}
\end{table}

The results follow the random matrix logic closely. With a sixty-day window ($q=0.82$), the sample estimator produces a realised volatility of \VolSampleSixty{}\%, worse than the \EWVol{}\% of the equally weighted portfolio, which is the noise-dressing failure documented by \citet{pafka2003noisy} and \citet{demiguel2009optimal}. Clipping and linear shrinkage cut that volatility in half, to \VolClipSixty{}\% and \VolLWSixty{}\%. The rotationally invariant estimator does less well at this extreme ratio (\VolRIESixty{}\%), because the sample Stieltjes transform that it relies on is poorly resolved with only 49 eigenvalues. As the window lengthens the estimators converge. At $T=250$ the spread between the best and worst is less than half a percentage point, and at $T=500$ all four lie between \VolSampleFiveHundred{}\% and \VolClipFiveHundred{}\%. At that window length clipping is slightly worse than the sample estimator, because it discards genuine sector structure below the edge. The realised-to-predicted variance ratios of the sample estimator exceed the Gaussian benchmark at every window. Examples are \RatioTwoFifty{} against \TheoryTwoFifty{} at $T=250$ and \RatioFiveHundred{} against \TheoryFiveHundred{} at $T=500$. The excess is the effect of non-stationarity and heavy tails, which inflate out-of-sample risk for every estimator. Shrinkage and the rotationally invariant estimator, which are not subject to the $(1-q)$ factors, also show realised-to-predicted ratios of about $1.5$ at $T=500$. Measured relative to that common factor, the excess of the sample estimator is about $1.23$ at $T=250$ and $1.09$ at $T=500$. This is below the Gaussian benchmark, so in these data the two sources of out-of-sample error overlap rather than compound multiplicatively.

\subsection*{Information linkage between carbon-intensive and financial industries}

The information analysis asks how much the daily returns of carbon-intensive industries reveal about the returns of financial industries. This is a reduced-form version of the transition-risk channel studied in climate finance \citep{battiston2017climate,engle2020hedging,bolton2021investors,giglio2021climate,pastor2022dissecting}. The first block contains the $p=\MIp$ industries Petroleum and Natural Gas, Coal, Utilities, Non-Metallic and Industrial Metal Mining, Steel Works, Chemicals, Transportation and Automobiles. The second contains the $r=\MIr$ industries Banking, Insurance, Real Estate and Trading. Within each 250-day window the returns are demeaned, so the relevant degrees of freedom are $n=\MIn$, and the plug-in information $\hat I$ is computed from the $12\times12$ sample covariance matrix. For this design the exact spurious information is $B_{\MIn}(\MIp,\MIr)=\MIBias$ nats, and the exact null standard deviation is $V_n^{1/2}=\MISd$. By the value-of-information theorem the same numbers, divided by $\gamma$, measure the certainty-equivalent value that an investor in the financial block would attach to observing the carbon block, were the latter available before trading. I report three variants. The first is the corrected information $\tilde I=\hat I-B_n$. The second is a placebo in which the carbon block is circularly shifted by a random offset of at least 21 days within each window, which destroys contemporaneous dependence while preserving each block's own dynamics and marginal distributions. The third is the partial information conditional on the equal-weighted market return, whose exact spurious information is $B_{n-1}(p,r)=\MIPartialBias$ nats because the regression uses one further degree of freedom.

\begin{figure}[htbp]
\centering
\includegraphics[width=\textwidth]{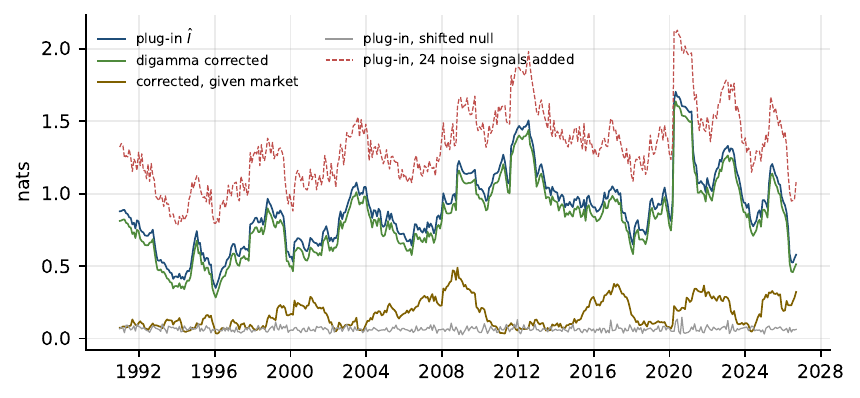}
\caption{Mutual information between eight carbon-intensive and four financial industries in 250-day rolling windows. Plug-in and digamma-corrected estimates, the corrected partial information given the equal-weighted market return, the plug-in estimate in a circularly shifted placebo, and the plug-in estimate after appending 24 independent Gaussian series to the financial block.}
\label{fig:rollingmi}
\end{figure}

Figure~\ref{fig:rollingmi} shows the results. The placebo is the cleanest test of the theory on real data. Its plug-in information averages \MINullPlugMean{} nats across windows against the exact prediction of \MIBias{}. Its corrected value averages \MINullCorrMean{}, and the standard deviation across windows is \MINullSd{}, against the exact $V_n^{1/2}=\MISd$. The one-sided corrected test rejects in \MINullRej\% of placebo windows at the nominal 5\% level. Despite heavy tails, volatility clustering and serial dependence, the Gaussian finite-sample theory of spurious information is accurate to the third decimal on these data. One reason is that the null law depends on the data only through the column spaces of the two blocks. It is exact whenever either column space is uniformly distributed, and in these data the departures from that condition caused by heavy tails and volatility clustering appear to be second order.

The genuine linkage is large and highly variable. The corrected information averages \MICorrMean{} nats, compared with \MIPlugMean{} for the plug-in estimate. It is significant at the 5\% level in every window (\MIShareSig\%), ranging from \MICorrMin{} nats in \MICorrMinDate{} to \MICorrMax{} nats in \MICorrMaxDate{}, and it has a correlation of \MICorrAbs{} with the absorption ratio. Most of the information that carbon-intensive returns carry about financial returns is therefore market-wide information. Conditioning on the market removes about four fifths of it. The median share of the partial information in the total is \MIPartialShare\%. The remainder is economically and statistically distinct. The corrected partial information averages \MIPartialMean{} nats, is significant at the 5\% level in every window (\MIPartialShareSig\%), even at its minimum of \MIPartialMin{} nats in \MIPartialMinDate{}, peaks at \MIPartialMax{} nats in \MIPartialMaxDate{}, and has a correlation of only \MIPartialAbs{} with absorption. The partial linkage has its own dynamics. It rises in the run-up to and during the financial crisis, again in 2016--2017 following the commodity downturn of 2015--2016, after 2020 and in 2025--2026. None of these episodes is marked by unusually high aggregate correlation. A sector-specific informational channel between carbon-intensive industries and finance is thus present throughout the sample and is not a by-product of the market mode.

Finally, the experiment with appended noise reproduces the monotonicity proposition on real data. Adding eight independent Gaussian series to the financial block raises the average plug-in information from \MIPlugMean{} to \MIPlugNoiseEight{} nats, and adding twenty-four raises it to \MIPlugNoiseTwentyFour{}. The corresponding exact spurious information is \MIBiasNoiseEight{} and \MIBiasNoiseTwentyFour{} nats, and the predicted increments match the observed increments to the second decimal. The corrected estimates remain at \MICorrNoiseEight{} and \MICorrNoiseTwentyFour{}, indistinguishable from the \MICorrMean{} of the original design. In value terms, an evaluator who did not correct would have attributed to the padded signal set an additional $0.42/\gamma$ of certainty-equivalent value. None of that value exists.
\section*{Discussion}

The results share a single mechanism. A Gaussian information functional is a difference of log-determinants, the log-determinant of a sample covariance matrix is a logarithmic linear spectral statistic, and the logarithm is concave. Each sample spectrum is spread around its population counterpart by the Marchenko--Pastur or Wachter law, so every log-determinant is biased downward, and by an amount that grows with dimension. Mutual information is a signed combination of three log-determinants whose dimensions sum to zero. The leading terms cancel, but the concavity does not, and what remains is the spurious information. Kullback--Leibler costs combine a log-determinant with a trace of an inverse and inherit the divergence of the latter at the left edge of the spectrum. The value of information is mutual information in other units. Random matrix theory thus provides not only the size of these effects but also their structure: exact digamma telescopes at finite $n$, Wachter integrals in the proportional limit, and phase transitions where low-rank structure meets the bulk.

Several qualifications apply. The exact results assume independent Gaussian observations, and financial returns are neither independent nor Gaussian. The placebo evidence in the application suggests that the null law is remarkably robust for block designs of moderate size. Robustness should not be assumed for heavier tails or larger dimensions, where the top of the spectrum can be dominated by a few extreme observations \citep{biroli2007top}. Elliptical and heavy-tailed versions of the theory, built on robust scatter estimators or on the general Silverstein equation with random scale, are natural extensions. The value-of-information identity is partial-equilibrium and relies on CARA preferences and joint normality. Under other preferences, mutual information remains an upper bound on the value of information for some classes of decision problems \citep{cabrales2013entropy}, but the equality is lost. The information frontier rests on the eigenvalue part of the canonical-correlation transition, which is established, and on a conjectured eigenvector counterpart, which this paper supports only by analogy with the BBP case. The dynamic analysis uses the simplest rotation-invariant model. Empirical correlation dynamics have a market mode that breaks rotation invariance, and the sufficiency result should then be applied to the bulk after projecting out the market direction.

The methodological implications are nonetheless direct. Wherever a high-dimensional Gaussian information measure is estimated, whether entropy, mutual information, transfer entropy or a Kullback--Leibler divergence between fitted Gaussian models, the exact finite-sample bias is available and should be removed. The correction is exact, needs no knowledge of the true covariance, and remains valid in the regime where the bias is largest relative to the signal. It also differs from covariance cleaning, which is the right tool for quadratic losses but biases information estimates towards zero. For empirical work on connectedness, spillovers and information transmission in finance, the corrected statistics and the exact null law give a principled way to separate genuine information from the mechanical contribution of dimension. For the economics of information, the identity between mutual information and certainty-equivalent value means that data that look valuable in high dimensions may owe much of their apparent value to dimensionality. The spurious component can be computed exactly, and it grows whenever uninformative series are added.

\section*{Conclusion}

This paper has developed the random matrix theory of Gaussian information functionals and taken it to financial data. The plug-in mutual information between two blocks of variables carries an exact spurious component that depends only on the dimensions and the sample size. It has an exact log-Beta law under independence, grows linearly in the sample size at the rate $\kappa(c_1,c_2)$ in the proportional regime, and increases strictly with every uninformative coordinate. Its spectral representation yields a closed form for the logarithmic moment of the Wachter law, and the canonical-correlation phase transition gives a frontier below which information cannot be located. An exact CARA--Gaussian identity turns all of these statements into statements about the certainty-equivalent value of information. For correlation dynamics, the Dyson eigenvalue process of a matrix Ornstein--Uhlenbeck market is informationally sufficient for its dissipation rate, which can be learned at a rate quadratic in the number of assets. On thirty-six years of US industry returns, the tools recover the market and sector modes with graded reliability, quantify the out-of-sample cost of noise dressing, and isolate an information linkage between carbon-intensive and financial industries that survives exact correction and conditioning on the market. The manuscript, data, reproduction code, simulation outputs and Lean proofs are available in the accompanying repository.

\FloatBarrier
\bibliographystyle{plainnat}
\bibliography{references}

\clearpage
\appendix
\section*{Appendix A. Proofs}
\addcontentsline{toc}{section}{Appendix A. Proofs}

Throughout, $\psi=\Gamma'/\Gamma$ is the digamma function and $\psi'$ the trigamma function. The facts used are that $\psi$ is strictly increasing and $\psi'$ strictly decreasing and positive on $(0,\infty)$, that $\psi(x)=\log x-1/(2x)+O(x^{-2})$ and $\psi'(x)=1/x+1/(2x^2)+O(x^{-3})$ as $x\to\infty$, and that for $\beta\sim\Beta(a,b)$ one has $\E\log\beta=\psi(a)-\psi(a+b)$ and $\Var\log\beta=\psi'(a)-\psi'(a+b)$. For $Y\sim\chi^2_k$, $\E\log Y=\psi(k/2)+\log2$ and $\Var\log Y=\psi'(k/2)$.

\subsection*{Log-determinants}

Let $W\sim\mathcal W_N(n,\Sigma)$ with $n\ge N$. Then $W=\Sigma^{1/2}W_0\Sigma^{1/2}$ with $W_0\sim\mathcal W_N(n,I)$, so $\log\det(W/n)-\log\det\Sigma=\log\det W_0-N\log n$. By the Bartlett decomposition, $W_0=LL^\top$ with $L$ lower triangular and $L_{ii}^2\sim\chi^2_{n-i+1}$ independent \citep[Ch.~3]{muirhead1982aspects}, so $\log\det W_0=\sum_i\log L_{ii}^2$. Taking expectations and variances gives \eqref{eq:logdet}. For the limit, write
\[
 \ell_n(N)=\sum_{i=1}^N\log\Bigl(1-\frac{i-1}n\Bigr)-\sum_{i=1}^N\frac1{n-i+1}+O\Bigl(\sum_{i\le N}(n-i+1)^{-2}\Bigr).
\]
The last term is $O(1/n)$ when $N/n\to q<1$. The second sum equals $-\log(1-q)+o(1)$. The trapezoidal rule applied to the first sum, whose summand $f(u)=\log(1-u)$ is smooth on $[0,q]$, gives $n\int_0^{N/n}\log(1-u)\,\dd u+\tfrac12[f(0)-f(N/n)]+O(1/n)$. Since $\int_0^q\log(1-u)\,\dd u=-(1-q)\log(1-q)-q=q\varphi(q)$, the result is $\ell_n(N)=N\varphi(N/n)+\tfrac12\log(1-N/n)+o(1)$. In particular $\ell_n(N)/N\to\varphi(q)$. The variance limit follows in the same way from $\psi'((n-i+1)/2)=2/(n-i+1)+O(n^{-2})$, which gives $\sum_{i\le N}2/(n-i+1)\to-2\log(1-q)$.

\subsection*{Exact spurious information}

If $nS\sim\mathcal W_{p+r}(n,\Sigma)$, then the diagonal blocks satisfy $nS_{11}\sim\mathcal W_p(n,\Sigma_{11})$ and $nS_{22}\sim\mathcal W_r(n,\Sigma_{22})$, because a principal submatrix of $X^\top X$ is the cross-product matrix of the corresponding columns of $X$. Hence, by \eqref{eq:logdet},
\[
 2(\E\hat I-I)=\ell_n(p)+\ell_n(r)-\ell_n(p+r)
 =\Psi_n(p)+\Psi_n(r)-\Psi_n(p+r)-(p+r-p-r)\log\frac n2,
\]
which is $2B_n(p,r)$. For the telescoping form, split $\Psi_n(p+r)=\Psi_n(p)+\sum_{i=1}^r\psi((n-p-i+1)/2)$, so that $2B_n=\sum_{i=1}^r[\psi((n-i+1)/2)-\psi((n-p-i+1)/2)]$. Each summand is strictly positive because $\psi$ is strictly increasing and $p\ge1$. Symmetry in $(p,r)$ is evident from the first form. The Lean theorems \texttt{twiceBias\_eq}, \texttt{twiceBias\_pos} and \texttt{twiceBias\_symm} prove the telescoping, positivity and symmetry for an arbitrary sequence $c_i$ in place of $\psi((n-i+1)/2)$, assuming only that the sequence is strictly antitone for positivity. The same statements with $n-1$ (demeaning) or $n-k$ (regression on $k$ common regressors) follow because the residual cross-product matrix is then $\mathcal W_{p+r}(n-k,\Sigma)$, and $I$ is unaffected by the normalisation of $S$.

\subsection*{Null law}

Under $\Sigma_{12}=0$, condition on the second data block $X_2$ and let $P_2$ be the projection onto its column space. Write $E=X_1^\top(I-P_2)X_1$ and $H=X_1^\top P_2X_1$. Given $X_2$, the rows of $X_1$ are independent $\N(0,\Sigma_{11})$, so $E\sim\mathcal W_p(n-r,\Sigma_{11})$ and $H\sim\mathcal W_p(r,\Sigma_{11})$ are independent, and their law does not depend on $X_2$. Since $\det S=\det S_{22}\det(S_{11}-S_{12}S_{22}^{-1}S_{21})$ and $n(S_{11}-S_{12}S_{22}^{-1}S_{21})=E$, $nS_{11}=E+H$, we obtain
\[
 \Lambda=\frac{\det S}{\det S_{11}\det S_{22}}=\frac{\det E}{\det(E+H)},
\]
which is Wilks' $\Lambda(p,n-r,r)$. Applying the Bartlett decomposition to $E+H$ and transforming shows that $\Lambda\stackrel d=\prod_{i=1}^p\beta_i$ with independent $\beta_i\sim\Beta((n-r-i+1)/2,r/2)$ \citep[Ch.~8]{anderson2003introduction}. The moments of $\hat I=-\tfrac12\log\Lambda$ follow from the log-Beta moments. Telescoping again shows that $\tfrac12\sum_{i\le p}[\psi((n-i+1)/2)-\psi((n-r-i+1)/2)]=B_n(p,r)$.

\subsection*{Proportional limits}

Let $p/n\to c_1$, $r/n\to c_2$, $c_1+c_2<1$. From the telescoping form and $\psi(x)-\psi(y)=\log(x/y)+O(1/\min(x,y))$,
\[
 \frac{2B_n(p,r)}n=\frac1n\sum_{i=1}^r\log\frac{n-i+1}{n-p-i+1}+O\Bigl(\frac1n\Bigr)\to\int_0^{c_2}\bigl[\log(1-u)-\log(1-c_1-u)\bigr]\dd u.
\]
With $\int_0^{c_2}\log(1-u)\,\dd u=-(1-c_2)\log(1-c_2)-c_2$ and $\int_0^{c_2}\log(1-c_1-u)\,\dd u=(1-c_1)\log(1-c_1)-(1-c_1-c_2)\log(1-c_1-c_2)-c_2$, the limit is $2\kappa(c_1,c_2)$. In the same way, $\psi'(x)=1/x+O(x^{-2})$ gives
\begin{multline*}
 V_n(p,r)=\frac12\sum_{i=1}^p\Bigl[\frac1{n-r-i+1}-\frac1{n-i+1}\Bigr]+O\Bigl(\frac1n\Bigr)\\
 \to\frac12\int_0^{c_1}\Bigl[\frac1{1-c_2-u}-\frac1{1-u}\Bigr]\dd u=\nu(c_1,c_2).
\end{multline*}
Under independence, $\hat I-B_n$ is a sum of $p$ independent centred variables $-\tfrac12(\log\beta_i-\E\log\beta_i)$. Each has variance $O(1/n)$ and absolute third moment $O(n^{-3/2})$, uniformly in $i$, because $\beta_i$ has both shape parameters of order $n$ and $\log\beta_i$ is approximately normal with variance $\psi'(a)-\psi'(a+b)\asymp1/n$. Lyapunov's condition holds with ratio $O(pn^{-3/2})/O(1)^{3/2}=O(n^{-1/2})$, which gives the central limit theorem. For general $\Sigma$, each of the three log-determinants in $\hat I$ has variance $O(1)$ by \eqref{eq:logdet} and \eqref{eq:phi}, so $\Var\hat I=O(1)$ by the Cauchy--Schwarz inequality. Hence $(\hat I-\E\hat I)/n\to0$ in $L^2$, and $\E\hat I/n=I/n+B_n/n\to I/n+\kappa$.

\subsection*{Logarithmic moment of the Wachter law}

Under independence the previous step gives $\hat I/n\to\kappa(c_1,c_2)$ in probability. On the other hand $\hat I/n=-\tfrac p{2n}\int\log(1-x)\,\hat F_p(\dd x)$. By \citet{wachter1980limiting}, $\hat F_p$ converges weakly almost surely to the law with density \eqref{eq:wachter}, which has no atoms when $c_1\le c_2$ and $c_1+c_2<1$. The largest squared sample canonical correlation converges almost surely to $\omega_+$ \citep{johnstone2008multivariate,bao2019canonical}, and $\omega_+<1$ because
\[
 1-\omega_+=(1-c_1)(1-c_2)+c_1c_2-2\sqrt{c_1c_2(1-c_1)(1-c_2)}=\Bigl(\sqrt{(1-c_1)(1-c_2)}-\sqrt{c_1c_2}\Bigr)^2>0,
\]
where the last inequality is strict because $(1-c_1)(1-c_2)-c_1c_2=1-c_1-c_2>0$. Therefore, with probability tending to one, all $\hat\rho_k^2$ lie in $[0,(1+\omega_+)/2]$, on which $\log(1-x)$ is bounded and continuous. Weak convergence then gives $\int\log(1-x)\,\hat F_p(\dd x)\to\int\log(1-x)\,w_{c_1,c_2}(x)\,\dd x$. Equating the two limits of $\hat I/n$ yields the theorem.

\subsection*{Information costs and portfolio ratios}

For the reverse divergence, $\E\tr(\Sigma^{-1}S)=N$ and $\E\log\det(\Sigma^{-1}S)=\ell_n(N)$. For the forward divergence, $\E S^{-1}=n\Sigma^{-1}/(n-N-1)$ for $n>N+1$ \citep[Ch.~3]{muirhead1982aspects}, so $\E\tr(S^{-1}\Sigma)=Nn/(n-N-1)$, and $\E\log\det(S^{-1}\Sigma)=-\ell_n(N)$. The limits follow from $\ell_n(N)/N\to\varphi(q)$ and $n/(n-N-1)\to(1-q)^{-1}$.

For the minimum-variance portfolio, whitening reduces the problem to $\Sigma=I$ with $a=\Sigma^{-1/2}\one$, and both ratios are invariant to the scale of $a$ and to rotations, so we may take $a=e_1$ and $W\sim\mathcal W_N(n,I)$, $S=W/n$. Then $R_{\rm in}/R=1/(n\,e_1^\top W^{-1}e_1)$ and $1/e_1^\top W^{-1}e_1=w_{11}-w_{12}^\top W_{22}^{-1}w_{21}\sim\chi^2_{n-N+1}$ \citep[Ch.~3]{muirhead1982aspects}, which gives $\E R_{\rm in}/R=(n-N+1)/n$. For the out-of-sample ratio, $R_{\rm out}/R=e_1^\top W^{-2}e_1/(e_1^\top W^{-1}e_1)^2$. With $V=W^{-1}$, the block inverse formula gives $V_{21}/V_{11}=-W_{22}^{-1}w_{21}$, so the ratio equals $1+\|W_{22}^{-1}w_{21}\|^2$. Given $W_{22}$, $w_{21}\sim\N(0,W_{22})$, so $\E[\|W_{22}^{-1}w_{21}\|^2\mid W_{22}]=\tr W_{22}^{-1}$. Since $W_{22}\sim\mathcal W_{N-1}(n,I)$, $\E\tr W_{22}^{-1}=(N-1)/(n-N)$, which gives $\E R_{\rm out}/R=1+(N-1)/(n-N)=(n-1)/(n-N)$.

\subsection*{Value of information}

Let $m=\E[v\mid s]$ and $\Sigma_{v|s}=\Var(v\mid s)$. Conditional on $s$, terminal wealth under the position $x$ is normal with mean $w_0+x^\top(m-p)$ and variance $x^\top\Sigma_{v|s}x$, so $\E[-e^{-\gamma w}\mid s]=-\exp\{-\gamma w_0-\gamma x^\top(m-p)+\tfrac{\gamma^2}2x^\top\Sigma_{v|s}x\}$. This is maximised at $x^*=(\gamma\Sigma_{v|s})^{-1}(m-p)$, with value $-\exp\{-\gamma w_0-\tfrac12(m-p)^\top\Sigma_{v|s}^{-1}(m-p)\}$. Ex ante, $y=m-p\sim\N(c,B)$ with $c=\mu-p$ and $B=\Var(m)=\Sigma_v-\Sigma_{v|s}$ by the law of total variance. For $A$ positive definite and $B$ positive semidefinite, completing the square gives
\[
 \E e^{-\frac12y^\top Ay}=\det(I+BA)^{-1/2}\exp\bigl\{-\tfrac12c^\top(A^{-1}+B)^{-1}c\bigr\}.
\]
To see this, assume first that $B$ is invertible. The exponent of the integrand is $-\tfrac12[y^\top Ay+(y-c)^\top B^{-1}(y-c)]$. Its Gaussian integral produces the factor $\det(B)^{-1/2}\det(A+B^{-1})^{-1/2}=\det(I+BA)^{-1/2}$. The residual quadratic form is $c^\top[B^{-1}-B^{-1}(A+B^{-1})^{-1}B^{-1}]c=c^\top(A^{-1}+B)^{-1}c$ by the Woodbury identity. The general case follows by continuity. With $A=\Sigma_{v|s}^{-1}$, $I+BA=\Sigma_v\Sigma_{v|s}^{-1}$ and $A^{-1}+B=\Sigma_v$, so
\[
 U_s=-e^{-\gamma w_0}\det(\Sigma_v\Sigma_{v|s}^{-1})^{-1/2}\exp\bigl\{-\tfrac12(\mu-p)^\top\Sigma_v^{-1}(\mu-p)\bigr\}=U_\varnothing\,e^{-I(v;s)},
\]
since $U_\varnothing$ is the same expression with $\Sigma_{v|s}=\Sigma_v$. Paying $c$ for the signal multiplies $U_s$ by $e^{\gamma c}$, so indifference holds at $c=I(v;s)/\gamma$. If $s$ is replaced by $Ts$ with $T$ invertible, then $\Sigma_{v|s}$ is unchanged, and so is the value.

\subsection*{Matrix Ornstein--Uhlenbeck information rate}

The coordinates $\{H_{ij}\}_{i\le j}$ of \eqref{eq:mou} are independent scalar Ornstein--Uhlenbeck processes with drift $-aH_{ij}$ and diffusion coefficients $N^{-1/2}$ (off-diagonal) and $(2/N)^{1/2}$ (diagonal). For linear drifts the Girsanov density is a true martingale on finite horizons \citep{sarkka2019applied}, and the relative entropy between the path laws with drifts $-aH$ and $-bH$, under the law with drift $-aH$, is
\[
 \KL=\frac12\E_a\int_0^T\Bigl[\sum_{i<j}\frac{(b-a)^2H_{ij}^2}{1/N}+\sum_i\frac{(b-a)^2H_{ii}^2}{2/N}\Bigr]\dd t
 =\frac N4(b-a)^2\int_0^T\E_a\tr H_t^2\,\dd t,
\]
because $\tr H^2=2\sum_{i<j}H_{ij}^2+\sum_iH_{ii}^2$. At stationarity, $\E H_{ij}^2=1/(2aN)$ for $i\ne j$ and $\E H_{ii}^2=1/(aN)$, so $\E\tr H^2=(N+1)/(2a)$, as certified by the Lean theorem \texttt{mou\_trace\_moment}. For the eigenvalue process, the Dyson system \eqref{eq:dyson} has non-colliding solutions for the orthogonal class, and the two regimes share the diffusion $\sqrt{2/N}$ and the repulsion term. The drift difference is $(b-a)\lambda_i$, so Girsanov gives $\tfrac12\E_a\int\sum_i(b-a)^2\lambda_i^2/(2/N)\,\dd t$, which is the same quantity because $\sum_i\lambda_i^2=\tr H^2$. The Lean theorems \texttt{mou\_rate\_closed}, \texttt{mou\_eigen\_closed} and \texttt{mou\_eigen\_sufficient} certify the algebra. The stationary laws of the two regimes are centred Gaussian on the $N(N+1)/2$ independent coordinates with variance ratio $b/a$, which gives the initial divergence $\tfrac{N(N+1)}4[b/a-1-\log(b/a)]$.

\subsection*{The forced Burgers equation}

Let $f_z(\lambda)=(z-\lambda)^{-1}$, so $g_t=N^{-1}\sum_if_z(\lambda_i)$ with $f_z'=(z-\lambda)^{-2}$ and $f_z''=2(z-\lambda)^{-3}$. It\^o's formula and \eqref{eq:dyson} give
\[
 \dd g_t=\frac1N\sum_if_z'(\lambda_i)\Bigl[\frac1N\sum_{j\ne i}\frac1{\lambda_i-\lambda_j}-a\lambda_i\Bigr]\dd t+\frac1{N^2}\sum_if_z''(\lambda_i)\,\dd t+\dd M_t,
\]
with a martingale $M$ whose quadratic variation is $O(N^{-2})$. Symmetrising the double sum and using
\[
 \frac{f_z'(\lambda_i)-f_z'(\lambda_j)}{\lambda_i-\lambda_j}=\frac1{(z-\lambda_i)(z-\lambda_j)^2}+\frac1{(z-\lambda_i)^2(z-\lambda_j)}
\]
gives $N^{-2}\sum_{i\ne j}(z-\lambda_i)^{-2}(z-\lambda_j)^{-1}=-g\,\partial_zg+O(N^{-1})$, since $\partial_zg=-N^{-1}\sum_i(z-\lambda_i)^{-2}$. The confinement term is $-aN^{-1}\sum_i\lambda_i(z-\lambda_i)^{-2}=a\,\partial_z(zg)$, because $\partial_z(zg)=N^{-1}\sum_i[(z-\lambda_i)^{-1}-z(z-\lambda_i)^{-2}]=-N^{-1}\sum_i\lambda_i(z-\lambda_i)^{-2}$. The It\^o term is $O(N^{-1})$. Letting $N\to\infty$ yields \eqref{eq:burgers}.
\section*{Appendix B. Lean Formalization}
\addcontentsline{toc}{section}{Appendix B. Lean Formalization}

The file \texttt{lean/RMTChecks.lean} in the replication repository contains 28 theorems that compile under Lean~4.29.0 with the standard distribution alone \citep{demoura2021lean}. It does not depend on mathlib, and it contains no \texttt{sorry} and no \texttt{native\_decide}. The file is checked with \texttt{lean RMTChecks.lean}. The purpose of the formalization is limited and precise. Lean certifies the combinatorial and algebraic steps that connect the analytic arguments of Appendix~A. It does not certify the measure-theoretic statements themselves, namely Wishart moments, the Bartlett decomposition, digamma asymptotics, weak convergence to the Wachter law and Girsanov's theorem. Table~\ref{tab:lean} lists the theorems by group.

\begin{table}[htbp]
\centering
\caption{Theorems in \texttt{lean/RMTChecks.lean}. ``Kernel computation'' means proved by \texttt{decide +kernel}, which evaluates both sides in the Lean kernel with exact integer arithmetic. ``Symbolic'' means proved for all values of the variables, by induction or by the \texttt{grind} decision procedure for commutative rings and fields.}
\label{tab:lean}
\small
\begin{tabular}{>{\raggedright}p{5.4cm}p{7.2cm}l}
\toprule
Theorem & Statement & Method\\
\midrule
\multicolumn{3}{l}{\emph{Moment combinatorics}}\\
\texttt{mp\_moments\_are\_narayana} & recursion \eqref{eq:mp-recursion} gives Narayana polynomials, $k\le10$ & kernel\\
\texttt{mp\_low\_moments} & $m_1,\ldots,m_4$ as quoted in the text & kernel\\
\texttt{catalan\_recursion\_matches} & Catalan recursion equals $\binom{2n}n/(n+1)$, $n\le12$ & kernel\\
\texttt{narayana\_rows\_sum\_}\newline\texttt{to\_catalan} & Narayana rows sum to Catalan numbers, $k\le10$ & kernel\\
\texttt{semicircle\_moments\_catalan} & Wigner moment recursion gives Catalan numbers & kernel\\
\midrule
\multicolumn{3}{l}{\emph{Spurious information}}\\
\texttt{sumTo\_add}, \newline\texttt{sumTo\_sub} & splitting and linearity of finite sums & symbolic\\
\texttt{sumTo\_nonneg}, \newline\texttt{sumTo\_pos} & sign of sums of nonnegative or positive terms & symbolic\\
\texttt{twiceBias\_eq} & telescoping form of \eqref{eq:spurious} & symbolic\\
\texttt{twiceBias\_nonneg}, \texttt{twiceBias\_pos} & positivity for antitone and strictly antitone sequences & symbolic\\
\texttt{twiceBias\_noise}, \texttt{twiceBias\_noise\_pos} & exact increment and strict growth in uninformative coordinates & symbolic\\
\texttt{twiceBias\_symm} & symmetry in $(p,r)$ & symbolic\\
\midrule
\multicolumn{3}{l}{\emph{Phase transitions}}\\
\texttt{mp\_edges} & sum and product of the edges $(1\pm\sqrt q)^2$ & symbolic\\
\texttt{bbp\_outlier\_at\_threshold} & $\lambda(1+\sqrt q)=(1+\sqrt q)^2$ & symbolic\\
\texttt{bbp\_overlap\_at\_threshold} & overlap numerator vanishes at threshold & symbolic\\
\texttt{bbp\_inverse} & outlier satisfies the inversion quadratic & symbolic\\
\texttt{cca\_outlier\_at\_threshold} & $\gamma(t_c)=\omega_+$ in \eqref{eq:cca-spike} & symbolic\\
\texttt{cca\_outlier\_at\_one} & $\gamma(1)=1$ & symbolic\\
\midrule
\multicolumn{3}{l}{\emph{Dynamics, costs and value}}\\
\texttt{mou\_trace\_moment} & $\E\tr H^2=(N+1)/(2a)$ from entry variances & symbolic\\
\texttt{mou\_rate\_closed}, \texttt{mou\_eigen\_closed} & entrywise and eigenvalue rates equal $N(N+1)(b-a)^2/(8a)$ & symbolic\\
\texttt{mou\_eigen\_sufficient} & equality of the two rates & symbolic\\
\texttt{stein\_trace} & $Nn/(n-N-1)-N=N(N+1)/(n-N-1)$ & symbolic\\
\texttt{gmv\_ratio} & product form of the portfolio ratios & symbolic\\
\texttt{cara\_scalar} & scalar determinant and inverse identities & symbolic\\
\bottomrule
\end{tabular}
\end{table}

The central statements about spurious information are proved for an arbitrary sequence $c:\mathbb N\to\mathbb Q$, which plays the role of $i\mapsto\psi((n-i)/2)$. The only property assumed is that the sequence is antitone, so the formal results apply to the digamma terms once the monotonicity of $\psi$ is granted. The two key statements read as follows in the source.

\begin{lstlisting}
def twiceBias (c : Nat → Rat) (p r : Nat) : Rat :=
  sumTo c p + sumTo c r - sumTo c (p + r)

theorem twiceBias_eq (c : Nat → Rat) (p r : Nat) :
    twiceBias c p r = sumTo (fun i => c i - c (p + i)) r

theorem twiceBias_noise_pos (c : Nat → Rat)
    (anti : ∀ i j, i < j → c j < c i) (p r m : Nat)
    (hp : 0 < p) (hm : 0 < m) :
    twiceBias c p r < twiceBias c p (r + m)
\end{lstlisting}

The information-rate result is stated with the stationary entry variances and diffusion coefficients written out explicitly, so that the closed form is checked against the same bookkeeping used in the Girsanov computation:

\begin{lstlisting}
def mouRateEntries (N a b : Rat) : Rat :=
  (1 / 2) * (b - a) ^ 2 *
    (N * (N - 1) / 2 * ((1 / (2 * a * N)) / (1 / N))
      + N * ((1 / (a * N)) / (2 / N)))

theorem mou_eigen_sufficient (N a b : Rat) (hN : N ≠ 0) (ha : a ≠ 0) :
    mouRateEigen N a b = mouRateEntries N a b
\end{lstlisting}
\section*{Appendix C. Additional Tables}
\addcontentsline{toc}{section}{Appendix C. Additional Tables}

Table~\ref{tab:practice} is a look-up table of exact spurious information for common designs. Each entry is $B_n(p,r)$ in nats, with the exact null standard deviation $V_n(p,r)^{1/2}$ in parentheses. When the data are demeaned, $n$ should be read as the sample size minus one, and when both blocks are regressed on $k$ common controls, as the sample size minus $k+1$. Corrected estimates are obtained by subtracting the entry from the plug-in mutual information, and the parenthesised value gives the scale of a one-sided test of zero information. Two features stand out. Spurious information is roughly proportional to $pr/n$ when both blocks are small, as the classical approximation predicts. It grows much faster once $(p+r)/n$ is appreciable. At $n=120$ and $(p,r)=(25,50)$ it reaches almost eight nats.

\begin{table}[htbp]
\centering
\caption{Exact spurious information $B_n(p,r)$ in nats, with null standard deviation $V_n(p,r)^{1/2}$ in parentheses, for block dimensions $(p,r)$ and degrees of freedom $n$. A dash indicates $n<p+r+2$.}
\label{tab:practice}
\footnotesize
\setlength{\tabcolsep}{4pt}
\begin{tabular}{rrrrrrrr}
\toprule
$n$ & $(2,2)$ & $(5,5)$ & $(5,20)$ & $(10,10)$ & $(10,40)$ & $(20,20)$ & $(25,50)$\\
\midrule
60 & 0.035 \scriptsize(0.025) & 0.230 \scriptsize(0.065) & 1.081 \scriptsize(0.154) & 1.017 \scriptsize(0.144) & 6.711 \scriptsize(0.516) & 5.304 \scriptsize(0.385) & --\\
120 & 0.017 \scriptsize(0.012) & 0.109 \scriptsize(0.031) & 0.469 \scriptsize(0.066) & 0.457 \scriptsize(0.065) & 2.151 \scriptsize(0.153) & 2.024 \scriptsize(0.144) & 7.945 \scriptsize(0.325)\\
250 & 0.008 \scriptsize(0.006) & 0.051 \scriptsize(0.014) & 0.211 \scriptsize(0.030) & 0.209 \scriptsize(0.030) & 0.893 \scriptsize(0.063) & 0.873 \scriptsize(0.062) & 2.965 \scriptsize(0.119)\\
500 & 0.004 \scriptsize(0.003) & 0.025 \scriptsize(0.007) & 0.103 \scriptsize(0.015) & 0.102 \scriptsize(0.014) & 0.422 \scriptsize(0.030) & 0.417 \scriptsize(0.030) & 1.354 \scriptsize(0.054)\\
1000 & 0.002 \scriptsize(0.001) & 0.013 \scriptsize(0.004) & 0.051 \scriptsize(0.007) & 0.051 \scriptsize(0.007) & 0.205 \scriptsize(0.015) & 0.204 \scriptsize(0.014) & 0.650 \scriptsize(0.026)\\
2500 & 0.001 \scriptsize(0.001) & 0.005 \scriptsize(0.001) & 0.020 \scriptsize(0.003) & 0.020 \scriptsize(0.003) & 0.081 \scriptsize(0.006) & 0.081 \scriptsize(0.006) & 0.254 \scriptsize(0.010)\\
\bottomrule
\end{tabular}
\end{table}

Table~\ref{tab:spurious} reports the numbers behind the left panel of Figure~\ref{fig:spurious}, and Table~\ref{tab:noise} those behind the right panel.

\begin{table}[htbp]
\centering
\caption{Null mean and standard deviation of the plug-in mutual information for independent blocks with $p=r$ at $n=250$: exact digamma values, proportional limit, classical approximation and simulation (\MCRepsQuarter{} draws per row).}
\label{tab:spurious}
\small
\begin{tabular}{rrrrrrr}
\toprule
$p=r$ & exact $\E_0\hat I$ & $n\kappa$ & $pr/(2n)$ & simulated & exact sd & simulated sd\\
\midrule
2 & 0.008 & 0.008 & 0.008 & 0.008 & 0.006 & 0.006 \\
5 & 0.051 & 0.051 & 0.050 & 0.051 & 0.014 & 0.014 \\
10 & 0.209 & 0.208 & 0.200 & 0.208 & 0.030 & 0.029 \\
20 & 0.873 & 0.871 & 0.800 & 0.874 & 0.062 & 0.061 \\
30 & 2.057 & 2.052 & 1.800 & 2.057 & 0.097 & 0.093 \\
40 & 3.842 & 3.833 & 3.200 & 3.845 & 0.136 & 0.135 \\
60 & 9.664 & 9.638 & 7.200 & 9.680 & 0.230 & 0.234 \\
80 & 19.651 & 19.588 & 12.800 & 19.655 & 0.355 & 0.358 \\
100 & 36.534 & 36.388 & 20.000 & 36.534 & 0.545 & 0.531 \\
\bottomrule
\end{tabular}
\end{table}

\begin{table}[htbp]
\centering
\caption{Mutual information estimates with $m$ appended uninformative signals ($n=250$, five payoffs, five informative signals with canonical correlation $0.4$), averaged over \MCRepsQuarter{} draws. ``Predicted'' is $I+B_n(5,5+m)$.}
\label{tab:noise}
\small
\begin{tabular}{rrrrrrr}
\toprule
$m$ & true $I$ & plug-in & predicted & corrected & Ledoit--Wolf & RIE\\
\midrule
0 & 0.436 & 0.485 & 0.487 & 0.434 & 0.283 & 0.397 \\
5 & 0.436 & 0.536 & 0.539 & 0.433 & 0.192 & 0.340 \\
10 & 0.436 & 0.595 & 0.593 & 0.438 & 0.130 & 0.297 \\
20 & 0.436 & 0.702 & 0.703 & 0.435 & 0.059 & 0.223 \\
40 & 0.436 & 0.937 & 0.939 & 0.434 & 0.016 & 0.148 \\
80 & 0.436 & 1.494 & 1.490 & 0.440 & 0.004 & 0.114 \\
\bottomrule
\end{tabular}
\end{table}

\end{document}